\documentclass[preprint, 10pt, 3p]{elsarticle}

\usepackage{amsmath,amsthm,amssymb,amsfonts,amsbsy,stmaryrd,latexsym}
\usepackage{epsfig}
\usepackage{framed}
\usepackage{epstopdf}
\usepackage{graphicx}
\usepackage{multirow}
\usepackage{subfigure}
\usepackage{tikz}
\usetikzlibrary{shapes,arrows,positioning}
\usepackage[ruled]{algorithm2e}

\usepackage{color, colortbl}
\usepackage{xcolor}
\usepackage{ulem}

\usepackage{subfigure}

\journal{Journal of Computational Physics}

\newcommand{\la}{\left\langle}
\newcommand{\ra}{\right\rangle}
\newcommand{\dd}[2]{ \frac{d {#1}}{d {#2} } }
\newcommand{\pd}[2]{ \frac{\partial {#1}}{\partial {#2}} }
\newcommand{\ul}[1]{\underline{{#1}}}

\newcommand{\hv}{\hat{v}}
\newcommand{\hw}{\hat{w}}

\newcommand{\bv}{\textbf{v}}
\newcommand{\bw}{\textbf{w}}
\newcommand{\hbv}{\hat{\textbf{v}}}
\newcommand{\hbw}{\hat{\textbf{w}}}

\newcommand{\bb}{{\bf b}}
\newcommand{\bg}{{\bf g}}
\newcommand{\bh}{{\bf h}}
\newcommand{\bA}{{\bf A}}
\newcommand{\bB}{{\bf B}}

\newcommand{\bU}{{\bf U}}
\newcommand{\bV}{{\bf V}}
\newcommand{\bR}{{\bf R}}

\newcommand{\ubv}{\underline{\bf v}}
\newcommand{\ubw}{\underline{\bf w}}
\newcommand{\uhbv}{\underline{\hat{\textbf{v}}}}
\newcommand{\uhbw}{\underline{\hat{\textbf{w}}}}
\newcommand{\ubb}{\underline{\bf b}}

\newcommand{\ubg}{\underline{\bf g}}
\newcommand{\ubx}{\underline{\bf x}}

\newcommand{\ubR}{\underline{\bf R}}

\newcommand*{\vertbar}{\rule[-1ex]{0.5pt}{2.5ex}}
\newcommand*{\horzbar}{\rule[.5ex]{2.5ex}{0.5pt}}

\begin{document}

\begin{frontmatter}

\title{Multiple Shooting Shadowing for Sensitivity Analysis of Chaotic Dynamical Systems}

\author[mit]{Patrick J. Blonigan\footnote{Current Address: NASA Ames Research Center, Moffett Field, CA 94035, United States}}
\ead{blonigan@mit.edu}

\author[mit]{Qiqi Wang}
\ead{qiqi@mit.edu}

\address[mit]{Department of Aeronautics and Astronautics,
Massachusetts Institute of Technology, 77 Massachusetts Ave, Cambridge,
MA 02139, United States}


\begin{abstract}
Sensitivity analysis methods are important tools for research and design with simulations. Many important simulations exhibit chaotic dynamics, including scale-resolving turbulent fluid flow simulations. Unfortunately, conventional sensitivity analysis methods are unable to compute useful gradient information for long-time-averaged quantities in chaotic dynamical systems. Sensitivity analysis with least squares shadowing (LSS) can compute useful gradient information for a number of chaotic systems, including simulations of chaotic vortex shedding and homogeneous isotropic turbulence. However, this gradient information comes at a very high computational cost. This paper presents multiple shooting shadowing (MSS), a more computationally efficient shadowing approach than the original LSS approach. Through an analysis of the convergence rate of MSS, it is shown that MSS can have lower memory usage and run time than LSS. 
\end{abstract}

\end{frontmatter}

\section{Introduction}

Computational methods for sensitivity analysis are invaluable tools for research and design in many engineering and scientific fields. These methods compute derivatives of outputs with respect to inputs in computer simulations. In applications with large amounts of parameters and only a few important outputs, adjoint-based sensitivity analysis is especially efficient \cite{Giles:2000:adj}. In aircraft design, for example, the number of geometric parameters that define the outer mold line is typically very large, but engineers may only be interested in a few outputs, such as the lift-to-drag ratio. As a result, the adjoint method of sensitivity analysis has proven to be very successful for aircraft design with gradient-based optimization \cite{Jameson:1988:adj,Reuther:2001:adjAC,Martins:2005:adjAS}. The adjoint method has also been an essential tool for adaptive grid methods for solving partial differential equations (PDE's) \cite{Darmofal:2002:adapt},  error estimation \cite{Giles:2002:adjEE}, and flow control problems \cite{Gunzburger:2002:PFC:640624}. Finally, some techniques for uncertainty quantification can benefit immensely from sensitivity information \cite{Wang:2009:Thesis,Wang:2013:hyper}.

Unfortunately, conventional sensitivity analysis methods, including the adjoint method, can break down when applied to chaotic systems. This occurs for sensitivities of long-time-averaged quantities of interest to design inputs. In this context ``long-time'' refers to time averaging horizons much larger than the physical time scales associated with the chaotic system being considered. This is problematic, as many key scientific and engineering quantities of interest in chaotic systems are long-time-averaged quantities, such as the time-averaged lift or drag coefficient of a flight vehicle in a high-lift configuration or the average heat transfer to a turbine blade. 

To carry out efficient design and analysis of chaotic systems and fluid flows, a new sensitivity analysis method is needed. One promising new approach is Least Squares Shadowing (LSS) \cite{Wang:2014:LSS2,Wang:2013:LSS_PF,Blonigan:2014:KS,Blonigan:2014:MG}. The most common implementation of LSS in the literature is called transcription LSS. Transcription LSS involves solving a globally coupled space-time problem, which can be computationally intensive \cite{Wang:2014:LSS2}. The study of LSS for chaotic vortex shedding by Blonigan et al. \cite{Blonigan:2016:AIAA} shows that transcription LSS is very costly in memory usage and operation count for a relatively small simulation. Similar issues with computational cost were encountered in a study of transcription LSS for a direct numerical simulation (DNS) of homogeneous isotropic turbulence \cite{gomez:2013:masters}. 

This paper presents an alternative way to pose the minimization statement to compute the shadowing direction or the adjoint shadowing direction. This formulation, multiple shooting shadowing (MSS), addresses the high computational cost of LSS. It is shown that MSS can reduce the memory requirements and the run time required to compute sensitivities of chaotic systems, making LSS sensitivity analysis more tractable for large chaotic dynamical systems such as turbulent fluid flow simulations. The convergence properties of MSS are presented in great detail, along with some approaches to control the convergence rate of MSS.

This paper is organized as follows: section \ref{s:past} discusses how conventional sensitivity analysis approaches break down for chaotic systems and the past work done to avoid this break down. Section \ref{s:lss} provides an overview of the formulation of transcription LSS. Section \ref{s:MSSform} introduces the MSS minimization statement and section \ref{s:mss_imp} discusses how MSS can be implemented. Next, section \ref{s:examples} shows MSS results for a chaotic dynamical system and a chaotic partial differential equation (PDE). Section \ref{s:mss_cond} discusses the convergence properties of MSS. Finally, section \ref{s:conclusion} summarizes this thesis and discusses some future research directions.

\section{Sensitivity Analysis of a Chaotic dynamical System}
\label{s:past}

The first question to be asked is if sensitivities are in fact well defined for chaotic systems. It is believed that many, but not every, chaotic system has differentiable time averaged quantities $\overline{J}$. Specifically, chaotic systems classified as uniformly hyperbolic or quasi-hyperbolic have differentiable time averaged quantities, but non-hyperbolic systems do not. These classes are discussed in greater detail below. 

A uniformly hyperbolic attractor is a strange attractor with a tangent space that can be decomposed into stable, neutrally stable, and unstable subspaces at every point in phase space \cite{Hasselblatt:2002:hyperbolic}. In other words, the Lyapunov covariant vectors make up a basis for phase space at all points on an attractor. Ruelle's linear response theorem states that hyperbolic attractors have mean quantities that respond differentiably to small perturbations to their parameters \cite{Ruelle:1997:SRB}. Therefore, sensitivities are well defined for chaotic systems with hyperbolic attractors. A well studied example of a hyperbolic attractor is the Plykin attractor \cite{Kuznetsov:2009:plykin}. The equations governing the Plykin attractor were designed to have hyperbolic properties, which are rare in practice. 

Although uniformly hyperbolic attractors are rare, many important properties of hyperbolic systems, including Ruelle's linear response theorem, can also be shown to hold for the far more common non-uniformly hyperbolic or quasi-hyperbolic attractors \cite{Ruelle:1997:SRB,Eyink:2004:ensmbl,Bonatti:2005:Hyper}. One example of a quasi-hyperbolic attractor is the Lorenz attractor \cite{Lorenz:1963:det}. At the origin of phase space, the Lyapunov covariant vectors for the positive and negative exponent are parallel, so hyperbolicity does not apply. However, this point is an unstable saddle point and almost all phase space trajectories do not pass through it. Because of this the Lorenz attractor appears to have the properties of a hyperbolic attractor, most importantly differentiable mean quantities \cite{Lea:2000:climate_sens,Wang:2014:LSS2}. 

Other chaotic dynamical systems have non-hyperbolic attractors. In these non-hyperbolic systems the time averaged quantities are usually not differentiable or even continuous as the parameters vary. In fact, long-time-averages for non-hyperbolic systems may have nontrivial dependence on the initial condition (i.e. the system is not ergodic), which leads to time averaged quantities that are not well-defined. 

Fortunately, Gallavotti and Cohen's chaotic hypothesis conjectures that larger systems behave more like hyperbolic systems than non-hyperbolic systems \cite{Gallavotti:1995:chaosHyp1,Gallavotti:1995:chaosHyp2}. That is, larger systems should have differentiable infinite time-averaged quantities. Additionally, a study by Albers and Sprott found that larger chaotic systems tend to have smoothly varying topology changes in the attractor as system parameters are varying \cite{Albers:2006:StrucStab}. Long-time-averaged quantities do not necessarily vary smoothly across sudden topology changes like bifurcations. Therefore, the chaotic hypothesis and the work by Albers and Sprott suggest there are well defined sensitivities to be computed for a large range of chaotic systems, especially if these systems have a large number of degrees of freedom (DoF). This is encouraging considering the large numbers of DoF's in simulations such as those of chaotic and turbulent fluid flows. 

Additionally, there is some evidence that the chaotic hypothesis applies to simulations of turbulent fluid flows. Grid convergence studies have been done in many cases to ensure that the discretization of the governing equations is sufficiently detailed. For example, Kim et al. used a direct numerical simulation (DNS) to compute turbulent statistics such as the mean velocity profile of a turbulent channel flow with a coarse and a fine spatial discretization \cite{Kim:1986:dns_channel} to check if their fine discretization was fine enough. The statistics were the same for the coarse and fine discretizations, which shows that long-time-averaged quantities of the DNS respond smoothly to perturbations in the spatial discretization, as predicted by the chaotic hypothesis. Similar results of grid convergence studies for other DNS and Large Eddy Simulation (LES) results \cite{Medic:2012:CTR,Bose:2010:expLES} also support the chaotic hypothesis, making it very likely that sensitivities of long-time-averaged quantities are well defined for high-fidelity turbulent flow simulations. 

\subsection{Conventional Sensitivity Analysis}
\label{ss:ivp}

Many time dependent simulations can be interpreted as dynamical systems:

\begin{equation}
\dd{u}{t} = f(u; s),
\label{e:dyn_sys}
\end{equation}

where $u$ is a length $n$ vector representing the system state and $s$ is a set of design parameters. For a computational fluid dynamics (CFD) simulation, equation \eqref{e:dyn_sys} represents the discretized Navier-Stokes equations and the state variable $u$ is a vector containing the conserved quantities. The design parameters $s$ could include geometric parameters for a wing or flow parameters such as the Reynolds number. 

When conducting design studies, engineers are typically interested in minimizing some objective function, $J$:

\[
J(t;s) = J(u(t;s))
\]

One example of $J$ is the instantaneous drag on an airfoil. For unsteady simulations, time-averaged objective functions are often of interest:

\[
\bar{J}(s) = \frac{1}{T_1-T_0} \int_{T_0}^{T_1} J(t;s)dt
\]

Sensitivity analysis seeks to compute the derivative of $\bar{J}$ with respect to the design parameters $s$. Traditionally, sensitivity analysis is conducted by solving the following initial value problems  

\begin{align*}
 \dd{u}{t} &= f(u;s), \quad u(t=T_0) = u_0 \\
 \dd{u'}{t} &= f(u';s + \delta s), \quad u'(t=T_0) = u_0
\end{align*}

\noindent where $u(t)$ is a reference solution and $u'(t)$ is solution corresponding to a perturbation $\delta s$ to some design parameter. Sensitivities can be estimated as:

\begin{equation}
 \dd{\bar{J}}{s} \approx \frac{1}{\Delta T} \int_{T_0}^{T_0+\Delta T} \frac{J(u') - J(u)}{\delta s}  \ dt
 \label{e:obj_fd}
\end{equation}

\noindent However, this initial value problem is very poorly conditioned when the system is chaotic \cite{Wang:2013:LSS_PF}. The ``butterfly effect'' ensures that $u(t)$ and $u'(t)$ will become decorrelated after some time. This is because when $u'(t)-u(t)$ is infinitesimal:

\begin{equation}
u'(t)-u(t) \sim e^{\Lambda^{max} t}
\label{e:lyapunov}
\end{equation}

\noindent where $\Lambda^{max}$ is the largest Lyapunov exponent of the system \cite{Strogatz:1994:chaos} (see \ref{a:lyapunov} for one way to compute this). Since chaotic dynamical systems have at least one positive Lyapunov exponent, $u'(t)-u(t)$ grows exponentially. Therefore, small perturbations to the initial condition or parameter $s$ will grow relatively large after a time $\mathcal{O}(1/\Lambda^{max})$. 

Equation \eqref{e:lyapunov} also implies that the linearization $v=\lim_{\delta s \to 0} \frac{u'(t)-u(t)}{\delta s}$ will grow exponentially. This growth of $v(t)$ is the cause of the following inequality

\begin{equation}
 \dd{\bar{J}}{s} = \lim_{\delta s \to 0} \lim_{\Delta T \to \infty}  \frac{1}{\Delta T} \int_{T_0}^{T_0+\Delta T} \frac{J(u') - J(u)}{\delta s}  \ dt \neq \lim_{\Delta T \to \infty}  \frac{1}{\Delta T} \int_{T_0}^{T_0+\Delta T} \lim_{\delta s \to 0} \frac{J(u') - J(u)}{\delta s}  \ dt 
 \label{e:obj_ineq}
\end{equation}

\noindent or

\[
 \dd{\bar{J}}{s} = \lim_{\delta s \to 0} \lim_{\Delta T \to \infty}  \frac{1}{\Delta T} \int_{T_0}^{T_0+\Delta T} \frac{J(u') - J(u)}{\delta s}  \ dt \neq \lim_{\Delta T \to \infty}  \frac{1}{\Delta T} \int_{T_0}^{T_0+\Delta T} \left(\pd{J}{u} v(t) + \pd{J}{s} \right) \ dt
\]

Therefore, sensitivity analysis formulated as an initial value problem does not compute useful sensitivities for chaotic dynamical systems. These same issues arise for conventional adjoint sensitivity analysis as well, as shown by Lea et al. for the Lorenz 63 equation \cite{Lea:2000:climate_sens}. 

\subsection{Chaotic Sensitivity Analysis}

Prior work in sensitivity analysis of long-time-averages in chaotic dynamical systems and fluid flows has been done mostly by the climatological and meteorological community. This work includes the ensemble-adjoint method proposed by Lea et al. \cite{Lea:2000:climate_sens}. This method has been applied to the Lorenz 63 system and an ocean circulation model \cite{Lea:2002:ocean}. Eyink et al. then went on to generalize the ensemble-adjoint method \cite{Eyink:2004:ensmbl}. The ensemble-adjoint method involves averaging over a large number of adjoint calculations for different segments of a time horizon (or time horizons). It was found by Eyink et al. that the sample mean of sensitivities computed with the ensemble adjoint approach for the Lorenz 63 system converges slower than $N^{-0.5}$, where $N$ is the number of samples, making it less computationally efficient than a naive Monte-Carlo approach \cite{Eyink:2004:ensmbl}. Recently, Ashley and Hicken explored using the ensemble adjoint approach for gradient-based optimization of a chaotic system, but encountered similar issues and observed similarly slow convergence as Eyink et al. \cite{Hicken:2014:chaosOpt}. 

Another idea is the Fokker-Planck adjoint approach for climate sensitivity analysis \cite{Thuburn:2005:FP}. This approach involves 
finding an invariant measure or stationary density which satisfies a Fokker-Planck equation 
to model the long-time-averaged dynamics, called the ``climate'' of the system in much of the literature. The adjoint of this Fokker-Planck equation is then used to compute derivatives with respect to long-time-averaged quantities. Fokker-Planck methods typically require discretizing phase space, which limits the method to fairly low dimensional systems. Also many variants of the method require adding diffusion into the system, potentially making the computed sensitivities inaccurate. A Fokker-Planck method proposed by the author and Wang eliminates the need for much of this additional diffusion, but requires a discretization of a manifold approximating the strange attractor, which poses challenges for higher-dimensional, less well understood strange attractors \cite{Blonigan:2014:density}. 

An analysis based on the Fokker-Planck equation produces the Fluctuation Dissipation Theorem (FDT) \cite{Nyquist:1928:therEq, Kubo:1966:FDT}. For conservative and nearly conservative dynamical systems, the FDT can be used to accurately compute climate sensitivities \cite{leith:1975:climate}. Several improved algorithms based on FDT have since been developed for computing climate sensitivity of non-conservative systems \cite{majda:2005:information, Abramov:2008:FDT,Abramov:2007:FDTblend}. However, for strongly dissipative systems whose SRB measure \cite{youngSRB} deviates strongly from Gaussian, FDT based methods can be inaccurate. Additionally, some of the proposed approaches require computing positive Lyapunov exponents and their corresponding covariant vectors, the current algorithm for which is prohibitively expensive for large systems \cite{Abramov:2008:FDT,Benettin:1980:Lyapunov}.

\section{Transcription Least Squares Shadowing}
\label{s:lss}  

The poor conditioning of the initial value problem in section \ref{ss:ivp} arises from the problem formulation: $u(t)$ and $u'(t)$ will only be close in phase space at $t=T_0$ where they share the initial condition $u_0$; nothing in the problem formulation requires $u(t)$ and $u'(t)$ to be correlated. LSS overcomes the issues of the initial value problem by minimizing the distance in phase space between $u(t)$ and $u'(t)$ in a least squares sense \cite{Wang:2014:LSS2}. This is done by assuming ergodicity and replacing the initial condition with a regularization, as in Doedel and Friedman's continuation method for computing heteroclinic orbits \cite{Doedel:1989:orbits}. This regularization forces $u'(t)$ and $u(t)$ to be as close to one another in phase space as possible for $T_0 \le t \le T_1$:

\begin{equation}
 \min_{u,\tau} \frac{1}{2}\int_{T_0}^{T_1} \| u'(\tau(t) - u(t) \|^2 + \alpha \|1-d\tau/dt \|^2 \ dt \qquad s.t. \quad \, \dd{u'}{\tau} = f(u';s+\delta s), \quad T_0 \le t \le T_1
 \label{e:LSSnl}
\end{equation}

\noindent where $\tau(t)=(1+\delta s \eta t)$ is a time transformation whose purpose is explained in other LSS literature, and $\eta$ called the time dilation term \cite{Wang:2014:LSS2}.  

LSS uses the assumption of ergodicity to convert an initial value problem to a boundary value problem to improve the conditioning of solving for $u'(t)$. The solution to this boundary value problem, $u'(t)$, called the ``shadow trajectory'', has its existence guaranteed by the shadowing lemma for uniformly hyperbolic systems \cite{Pilyugin:1999:shadow,Wang:2014:LSSthm,Wang:2014:LSS2}. 

To compute sensitivities efficiently, equation \eqref{e:LSSnl} is linearized \cite{Wang:2014:LSS2}:

\begin{equation}
 \min_{v,\eta} \frac{1}{2} \int_{T_0}^{T_1} \| v(t) \|^2 + \alpha^2 \| \eta \|^2 \ dt, \quad \text{s.t.} \ \frac{dv}{dt} = \frac{\partial f}{\partial u} v + \frac{\partial f}{\partial s} + \eta f, \quad T_0 \le t \le T_1
 \label{e:opt_problem}
\end{equation}

\noindent where $v(t) = \lim_{\delta s \to 0} \frac{u'-u}{\delta s}$ is called the ``shadowing direction''. Equation \eqref{e:opt_problem} is a linearly constrained least-squares problem with the following Karush-Kuhn-Tucker (KKT) conditions derived using calculus of variations:

\begin{align}
\frac{\partial w}{\partial t} &= -\left(\frac{\partial f}{\partial u}\right)^* w - v, \quad
w(0) = w(T) = 0 \label{e:KKTw}\\
&\alpha^2 \eta = \langle f, w \rangle \label{e:KKTeta}\\
\frac{dv}{dt} &= \frac{\partial f}{\partial u} v + \frac{\partial f}{\partial s} + \eta f \label{e:KKTv}
\end{align}

To compute many sensitivities efficiently, one can derive the adjoint equations for equation \eqref{e:opt_problem}. Adjoint LSS is derived by Wang et al. \cite{Wang:2014:LSS2}.  

The main approach used to numerically solve for the shadowing direction (or the adjoint shadowing direction) is called transcription LSS. Transcription LSS is when equations \eqref{e:KKTw}, \eqref{e:KKTeta}, and \eqref{e:KKTv} are discretized and solved simultaneously for all time $T_0<t<T_1$ \cite{Wang:2014:LSS2}. The KKT system formed by equations \eqref{e:KKTw}, \eqref{e:KKTeta}, and \eqref{e:KKTv} can be reduced to a $mn \times mn$ block tridiagonal linear matrix, where $m$ is the number of time steps and $n$ is the number of degrees of freedom of the system. The KKT system is solved directly for smaller systems and solved using a multigrid in space and time scheme for larger systems \cite{Blonigan:2014:MG,gomez:2013:masters}. 

Transcription LSS has been used to compute accurate sensitivities for a computational fluid dynamics (CFD) simulation of chaotic vortex shedding \cite{Blonigan:2016:AIAA} and a DNS of homogeneous isotropic turbulence \cite{gomez:2013:masters}. Unfortunately, these studies also show that transcription LSS is very costly in memory usage and operation count. Much of this cost is due to the large size of the KKT system. For example, the chaotic vortex shedding simulation has $n=11,090$ degrees of freedom. For the $m=2,000$ time step window considered in the study, the total size of the KKT Schur complement matrix is 22,180,000 rows and roughly 24 GB of input data is required \cite{Blonigan:2016:AIAA}. The system was solved on 2,000 cores for roughly 17 hours to converge the sensitivity to at least 3 decimal places. This high cost provides motivation for the work presented in this paper on the multiple shooting implementation of LSS, called Multiple Shooting Shadowing (MSS). 

\section{Multiple Shooting Shadowing Formulation}
\label{s:MSSform}

The cost of LSS can be reduced by reformulating the least squares minimization problem \eqref{e:opt_problem}. More specifically, the objective function in equation \eqref{e:opt_problem} can be modified to reduce the number of constraint equations needed to compute a shadowing direction $v(t)$. Instead of minimizing the integral of $v(t)$ over a time horizon as in transcription LSS \eqref{e:opt_problem}, MSS seeks to minimize $v(t)$ at discrete checkpoints in time. 

\begin{gather}
\min_{v(t_i)} \sum_{i = 0}^K \|v(t_i^+)\|_2^2 \label{e:MSSobj00} \\
\text{s.t.} \quad v(t_i^+) = v(t_i^-), \qquad i = 1,2,...,K \label{e:MSSconstraint00}\\
\frac{dv}{dt} = \frac{\partial f}{\partial u} v + \frac{\partial f}{\partial s} + \eta f  \,, \quad t_0<t<t_K \label{e:MSStangent}\\
\la f(u(t);s), v(t) \ra =0  \label{e:MSSdilation_cond}
\end{gather}

\begin{figure}
\centering
\includegraphics[width = 0.65\textwidth]{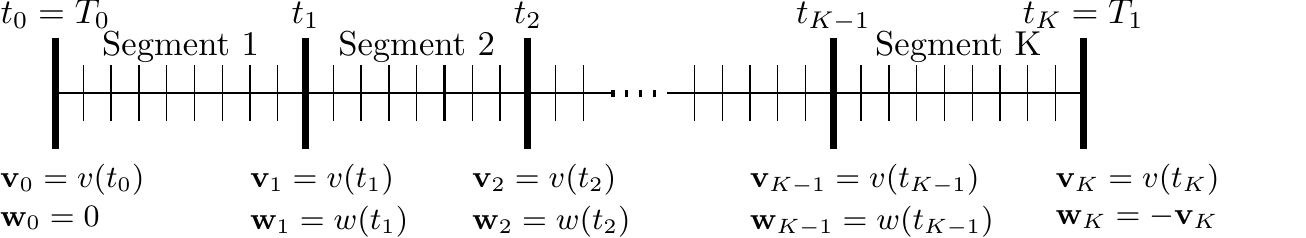}
\caption{Diagram showing the checkpoints and time segments used in MSS.  The boundary condition $\bw_K = -\bv_K$ follows from \eqref{e:MSSw}. }
\label{f:time_segments}
\end{figure}

Define time segment $i$ as the time span $t_{i-1} \le t \le t_i$, as shown in figure \ref{f:time_segments}. The MSS minimization problem penalizes large values of $v(t)$ at the end of each time segment. By doing this MSS seeks to find $v(t)$ that does not grow exponentially over each time segment. 

Equation \eqref{e:MSSconstraint00} shows that MSS has $Kn$ constraint equations for an $n$-DoF system, compared to $mn$ for transcription LSS, where $m$ is the number of time steps used to discretize the time horizon being studied. Since $K$ can be smaller than $m$, the MSS minimization problem can be smaller than that for transcription LSS. In fact, the number of time segments $K$ will almost always be smaller than then total number of time steps $m$. This is because the values of $K$ and $m$ are limited in different ways. If a time segment is too large, $v(t)$ will grow too large between checkpoints and round-off errors will impact the accuracy of MSS. The growth rate of $v(t)$ is at most the largest Lyapunov exponent $\Lambda^{max}$. This means that $\Lambda^{max}$ should be used to determine the minimum value of $K$ for a fixed time horizon. The number of time steps, $m$ is fixed by the desired temporal accuracy and the stability requirements of the time stepping scheme being used. Typically, the time step size needed for stability and accuracy is less than the time needed for round-off errors in the tangent solution to grow to $\mathcal{O}(1)$, so $K<m$. In fact, $K$ can be much smaller than $m$ when the discretization of the governing equation \eqref{e:dyn_sys} is stiff or the time accuracy requirements are strict. For example, for the Dowell's plate model presented later in this chapter, $m/K = 500$ for a relatively large choice of $K$. 

The constraint equation \eqref{e:MSSconstraint00} enforces continuity at each checkpoint, which forces $v(t)$ to satisfy the tangent governing equation \eqref{e:MSStangent}. Equation \eqref{e:MSSdilation_cond} fixes $\eta(t)$ to eliminate any component of $v(t)$ parallel to $f(u(t);s)$. In other words, equations \eqref{e:MSStangent} and \eqref{e:MSSdilation_cond} combine to form a well-defined algebraic-differential equation, in which $v$ contains the differential variables and $\eta$ is the algebraic variable.

Before proceeding to derive an expression for $\eta$, it is useful to define the {\it tangent propagator} $\phi^{t,t'}$

\begin{equation}
\phi^{t,t} = I \mbox{ for all } t\;,\quad
\phi^{t',\tau}\cdot\phi^{t,t'} = \phi^{t,\tau}\;,\quad
\frac{d}{d\tau} \phi^{t,\tau} = \frac{\partial f}{\partial u}\bigg|_{\tau}
\phi^{t,\tau}\;,\quad
\frac{d}{dt} \phi^{*\,t,\tau} = -\frac{\partial f}{\partial u}\bigg|_{t}^*
\phi^{*\,t,\tau}\;,
\label{e:tan_prop}
\end{equation}

\noindent where the superscript $*$ indicates an adjoint or transpose operator. The tangent propagator can be used to write the solution to equations \eqref{e:MSStangent} as

\begin{equation}
v(t) = \phi^{t_i,t} v(t_i)
+ \left(\int_{t_i}^{t} \eta_{i}(\tau)\,d\tau\right)f(u(t);s)
+ \int_{t_i}^{t} \phi^{\tau,t} \frac{\partial f}{\partial s}\bigg|_{\tau}\,d\tau\;,
\quad t_i\le t<t_{i+1}\;,
\label{e:tangent}
\end{equation}

The tangent solution $v(t)$ from equation \eqref{e:tangent} will satisfy \eqref{e:MSStangent} for any $\eta(t)$. For $v(t)$ to also satisfy equation \eqref{e:MSSdilation_cond}, $\int_{t_i}^{t} \eta_{i}(\tau)\,d\tau$ must satisfy the following closed form expression derived from equations \eqref{e:tangent} and \eqref{e:MSSdilation_cond}

\begin{equation}
 \int_{t_i}^{t} \eta_{i}(\tau)\,d\tau = - \frac{ \left\langle v'(t), f(u(t);s) \right\rangle}{\|f(u(t);s)\|^2_2}
 \label{e:v_proj0}
\end{equation}

\noindent where 

\[
v'(t) = \phi^{t_i,t} \bv_{i} + \int_{t_{i-1}}^{t_i} \phi^{\tau,t} \frac{\partial f}{\partial s}\bigg|_{\tau}\,d\tau
\]

\noindent Equation \eqref{e:v_proj0} and the above definition of $v'(t)$ can then be substituted into equation \eqref{e:tangent} to eliminate $\eta(t)$

\[
v(t) = v'(t) - \frac{ \left\langle v'(t), f(u(t);s) \right\rangle}{\|f(u(t);s)\|^2_2} f(u(t);s)
\quad t_i\le t<t_{i+1}\;,
\]

\noindent Therefore, the tangent solution $v(t)$ can be made to satisfy 
\eqref{e:MSSdilation_cond} by using the projection operator $P_t$

\begin{equation}
v(t) = P_t v'(t) \equiv v'(t) -  \frac{\langle f(u(t);s), v'(t) \rangle}{\|f(u(t);s)\|^2_2} f(u(t);s)
\label{e:v_proj}
\end{equation}

To derive the MSS KKT system, equations \eqref{e:tangent} and \eqref{e:v_proj} are substituted into equations \eqref{e:MSSobj00} and\eqref{e:MSSconstraint00}:

\begin{gather}
\min_{\bv_i} \sum_{i = 0}^K \|\bv_i\|_2^2 \label{e:MSSobj1} \\
\text{s.t.} \quad \bv_{i+1} = P_{t_{i+1}} \phi^{t_i,t_{i+1}} {\bf v}_{i}
+ P_{t_{i+1}} \int_{t_i}^{t_{i+1}} \phi^{\tau,t} \frac{\partial f}{\partial s}\bigg|_{\tau}\,d\tau 
\label{e:MSSconstraint1}
\end{gather}

\noindent where $\bv_i=v(t_i^+)$. 

For a dynamical system with $n$ degrees of freedom, \eqref{e:MSSconstraint1} can be written as

\begin{equation}
\bv_{i+1} = \Phi_{i+1} \bv_{i}
- \bb_{i+1}
\label{e:MSS0fwd}
\end{equation}

\noindent where $\Phi_{i+1} \equiv P_{t_{i+1}} \phi^{t_i,t_{i+1}}$ is an $n\times n$ matrix called the tangent transition matrix and 

\begin{equation}
\bb_i = -P_{t_i} \int_{t_{i-1}}^{t_i} \phi^{\tau,t} \frac{\partial f}{\partial s}\bigg|_{\tau}\,d\tau
\label{e:b_def}
\end{equation}

\noindent is a $n \times 1$ vector. 

All $K$ discretized constraints \eqref{e:MSS0fwd} can be written as a system of equations:

\begin{equation}
\uuline{\bB} \ubv = \ubb
\label{e:MSS0consMat}
\end{equation}

\noindent where

\[
\uuline{\bB} = \left(\begin{array}{ccccc}
\Phi_1 & -I & & & \\
 & \Phi_2 & -I & & \\
 & & \ddots & \ddots & \\
 & & & \Phi_K & -I 
\end{array}\right), \: \ul{\bf v} = \left(\begin{array}{c}
{\bf v}_0 \\
{\bf v}_1 \\
\vdots \\
{\bf v}_K
\end{array}\right), \: \ul{\bf b} = \left(\begin{array}{c}
{\bf b}_1 \\
{\bf b}_2 \\
\vdots \\
{\bf b}_K
\end{array}\right)
\]

The block matrix $\uuline{\bB}$ is $Kn$ by $K(n+1)$. The vectors $\ubv$, $\ubw$, and $\ubb$ are length $(K+1)n$, $Kn$, and $Kn$, respectively. 

 Using this notation, equation \eqref{e:MSSobj00} can be rewritten as

\begin{equation}
Q \equiv \sum_{i=0}^K \|\bv_i\|_2^2 = \ubv^T \ubv 
\label{e:MSS0objVec}
\end{equation}

Next a Lagrangian is formed for equations \eqref{e:MSS0objVec} and \eqref{e:MSS0consMat}. 

\[
\Lambda = \ubv^T \ubv + \ubw^T (\ubb -\uuline{\bB} \ubv)
\]

\noindent This Lagrangian is differentiated with respect to $\ubv$ and the Lagrange multiplier $\ubw$ to obtain the KKT equations

\begin{align*}
\pd{\Lambda}{\ubv} =& 0 = 2 \ubv^T - \ubw^T \uuline{\bB} \\
\pd{\Lambda}{\ubw} =& 0 = \ubb -\uuline{\bB} \ubv
\end{align*}

Together, the KKT equations form a system of equations that can be solved for $\ubv$ and $\ubw$:

\begin{equation}
\left(\begin{array}{c|c}
-\uuline{I} & \uuline{\bB}^T \\\hline 
\uuline{\bB} & 0 \end{array}\right) \left(\begin{array}{c}
\ubv\\\hline
\ubw
\end{array}\right) = \left(\begin{array}{c}
0 \\\hline
\ubb
\end{array}\right)
\label{e:MSS0sys}
\end{equation}

As for transcription LSS, the Schur complement of equation  \eqref{e:MSS0sys} is solved instead of the larger full KKT system. The KKT Schur complement for tangent MSS is

\begin{equation}
\uuline{\bB}\uuline{\bB}^T \ubw = \ubb
\label{e:MSS0schur1}
\end{equation}

\noindent or

\begin{equation}
\left(\begin{array}{cccc}
\Phi_1 \Phi_1^T + I & -\Phi_2^T & & \\
-\Phi_2 & \Phi_2 \Phi_2^T + I & -\Phi_3^T & \\
 & \ddots & \ddots & \ddots  \\
 & & -\Phi_K & \Phi_K\Phi_K^T + I 
\end{array}\right) \left(\begin{array}{c}
{\bf w}_1 \\
{\bf w}_2 \\
\vdots \\
{\bf w}_K
\end{array}\right) = \left(\begin{array}{c}
{\bf b}_1 \\
{\bf b}_2 \\
\vdots \\
{\bf b}_K
\end{array}\right)
\label{e:MSS0schur2}
\end{equation}

In practice, the matrix $\uuline{\bB}\uuline{\bB}^T$ is never formed, since each $\Phi_i$ is a dense $n \times n$ matrix and is expensive to compute. Instead, a routine for the product $\uline{\bf y} =\uuline{\bB}\uuline{\bB}^T \ubx$ is implemented and an iterative method is used to solve \eqref{e:MSS0schur1}, as shown in section \ref{s:mss_imp}. 

Although the matrix $\uuline{\bB}\uuline{\bB}^T$ is never formed, its structure reveals some key attributes of MSS. From equations \eqref{e:MSS0schur1} and \eqref{e:MSS0schur2}, it can be seen that the Schur complement matrix is a $Kn \times Kn$, symmetric positive definite and block tridiagonal matrix. As stated previously, $K$ is typically much smaller than the total number of time steps $m$. Therefore, the $Kn \times Kn$ MSS KKT Schur complement is potentially much smaller than the $mn \times mn$ KKT Schur complement for transcription LSS. 

Additionally, the structure of the MSS KKT Schur complement is independent of the time discretization used. This is different from transcription LSS, for which the structure of the KKT matrix system depends on the time discretization used \cite{Wang:2014:LSS2}. The second order Crank-Nicolson scheme considered by Wang et al. \cite{Wang:2014:LSS2} leads to a tridiagonal KKT Schur complement, but higher order time discretizations will lead to more bands in the KKT Schur complement matrix. Therefore MSS has a smaller KKT matrix and possibly also a smaller bandwidth than transcription LSS for a given time horizon and time discretization. This can lead to gains in computational efficiency and a potentially simpler implementation, as shown in section \ref{s:mss_imp}. 

It should be noted that as $K \to \infty$, the MSS minimization problem with uniform time segments becomes equivalent to the transcription LSS minimization problem \eqref{e:opt_problem} for $\alpha=0$, where $\alpha$ is the weighting parameter in \eqref{e:opt_problem}. This is explained in more detail in appendix \ref{a:MSS2tLSS}.

It should be also be noted that the block structure of equation \eqref{e:MSS0schur1} is similar to the LSS KKT system derived for maps \cite{Wang:2014:LSSthm}. This is because the operator $\Phi_i$ can be thought of as a Jacobian of a nonlinear mapping of $u$ from $t_i$ to $t_{i+1}$. For MSS, the nonlinear mapping is solving for $u(t_{i+1})$ from $u(t_i)$.  

As for transcription LSS, the sensitivity of a time-averaged objective function $\bar{J}$ is computed using the following expression \cite{Wang:2014:LSS2}

\[
\frac{d\bar{J}}{ds} = 
\frac1T \int_{t_0}^{t_K} \left( \left\langle \frac{\partial J}{\partial u}\bigg|_{t}, v \right\rangle + \eta \left(J - \bar{J} \right) \right) dt + \pd{\bar{J}}{s}
\]

\noindent for an averaging window from $t_0$ to $t_K$ and $T \equiv t_K-t_0$. The sensitivity of $\bar{J}$ to $s$ can be written as a function of $v'(t)$ from equation \eqref{e:v_proj} (see appendix \ref{a:MSSgrad} for the derivation of the second term)

\begin{equation}
\frac{d\bar{J}}{ds} =  
\frac1T \sum_{i=0}^{K-1}
 \int_{t_i}^{t_{i+1}} \left\langle \frac{\partial J}{\partial u}\bigg|_{t}, v' \right\rangle \ dt
+ \frac1T \sum_{i=0}^{K-1} \frac{\langle f_{i+1}, v'(t_{i+1}) \rangle}{\| f_{i+1} \|^2_2} \left(\bar{J} - J_{i+1}\right) + \pd{\bar{J}}{s}
\label{e:MSSgrad}
\end{equation} 

\noindent where $f_{i+1} = f(u(t_{i+1});s)$ and $J_{i+1} = J(u(t_{i+1});s)$. 

\subsection{The Filtering Parameter}

The MSS minimization problem expressed by equation \eqref{e:MSSobj00} and \eqref{e:MSSconstraint00} can be modified as follows for a better conditioned KKT system 

\begin{gather}
\min_{\bv_i,\bw_i} \sum_{i = 0}^K \left( \|\bv_i\|_2^2 + \epsilon\|\bw_i\|_2^2 \right) \label{e:MSSobj} \\
\text{s.t.} \quad \bv_{i} = v(t_i^-) + \epsilon \bw_{i}
\label{e:MSSconstraint}
\end{gather}

\noindent where $\bv_i = v(t_i^+)$ and $\bw_i = w(t_i^-)$, as shown in figure \ref{f:time_segments}, and $\epsilon$ is called the filtering parameter. Equation \eqref{e:MSSconstraint} shows that non-zero values of $\epsilon$, create a discontinuous jump of $\epsilon \bw_{i}$ in $v(t)$ at the checkpoints $t_i$. In other words, non-zero values of $\epsilon$ relax the continuity constraint \eqref{e:MSSconstraint00}. This has a major impact on the computational efficiency of MSS that will discussed in more detail in section \ref{s:mss_cond}. 

To derive the KKT equation for MSS with a non-zero $\epsilon$, start by rewriting equations \eqref{e:MSSobj} in the notation of equation \eqref{e:MSS0consMat}

\begin{equation}
Q \equiv \sum_{i=0}^K \left( \|\bv_i\|_2^2 + \epsilon\|\bw_i\|_2^2 \right) = \ubv^T \ubv + \epsilon \ubw^T \ubw 
\label{e:MSSobjVec}
\end{equation}

\noindent and \eqref{e:MSSconstraint} as

\begin{equation}
\uuline{\bB} \ubv + \epsilon \ubw = \ubb
\label{e:MSSconsMat}
\end{equation}

\noindent where $\ubw$ is a length-$Kn$ vector

\[
\ul{\bf w} = \left(\begin{array}{c}
{\bf w}_1 \\
{\bf w}_2 \\
\vdots \\
{\bf w}_K
\end{array}\right), 
\]

\noindent The constraint equation \eqref{e:MSSconsMat} can be used to eliminate $\ubw$ from equation \eqref{e:MSSobjVec}

\begin{equation}
Q = \ubv^T \ubv + \frac{1}{\epsilon} (\ubb - \uuline{\bB} \ubv)^T (\ubb - \uuline{\bB} \ubv)
\label{e:MSSobjVec2}
\end{equation}

\noindent As $Q$ is a quadratic function, it is convex, so the minimum of $Q$ satisfies the first order optimality condition:

\[
\pd{Q}{\ubv} = 0 = 2\ubv^T - \frac{2}{\epsilon} (\ubb - \uuline{\bB} \ubv)^T \uuline{\bB} 
\]

\noindent Using equation \eqref{e:MSSconsMat}, this expression can be rewritten as

\begin{equation}
-\ubv + \uuline{\bB}^T \ubw = 0
\label{e:MSSw}
\end{equation}

\noindent Together, equations \eqref{e:MSSconsMat} and \eqref{e:MSSw} form a system of equations that can be solved for $\ubv$ and $\ubw$:

\begin{equation}
\left(\begin{array}{c|c}
-\uuline{I} & \uuline{\bB}^T \\\hline 
\uuline{\bB} & \epsilon \uuline{I} \end{array}\right) \left(\begin{array}{c}
\ubv\\\hline
\ubw
\end{array}\right) = \left(\begin{array}{c}
0 \\\hline
\ubb
\end{array}\right)
\label{e:MSSsys}
\end{equation}

Finally, the KKT Schur complement for tangent MSS with non-zero $\epsilon$ is

\begin{equation}
\uuline{\bA} \ubw = \ubb, \qquad \uuline{\bA} =\uuline{\bB}\uuline{\bB}^T + \epsilon \uuline{I}, 
\label{e:MSSschur1}
\end{equation}

\noindent Note that the left-hand side is identical to that of equation \eqref{e:MSS0schur1} with $\epsilon$ times the identity matrix $\uuline{I}$ added on. In section \ref{s:mss_cond} it is shown that this additional term improves the conditioning of \eqref{e:MSSschur1}, allowing iterative solvers to converge to a solution in less time as $\epsilon$ is increased.


%
%
%
%
%

\subsection{Adjoint Formulation}

When the sensitivities of one time-averaged objective function $\bar{J}$ to many design parameters $s$ are desired, adjoint MSS can be used. Adjoint MSS is the discretely consistent adjoint of tangent MSS and can be expressed as the solution of the following linear system, (see appendix \ref{a:AdjMSS}):

\begin{equation}
\uuline{\bA} \uhbw =  -\uuline{\bB}\ubg, \qquad \uuline{\bA} =\uuline{\bB}\uuline{\bB}^T + \epsilon \uuline{I}, 
\label{e:MSSadjSchur1}
\end{equation}

\noindent where $\ubg^T = (\bg_1^T, \bg_2^T, ... \bg_K^T, 0)$ is a $1 \times (K+1)n$ vector and 

\begin{equation}
\bg_i^T =  
\frac1T \int_{t_{i-1}}^{t_{i}} \frac{\partial J}{\partial u}\bigg|_{t}^* \phi^{t_{i-1},t}  \ dt
+ \frac1T \left(\bar{J} - J_i\right) \frac{f_{i}|^* \phi^{t_{i-1},t_i}}{\|f_{i}\|^2_2} ,
\label{e:g_def1}
\end{equation}

The adjoint equations \eqref{e:MSSadjSchur1} and \eqref{e:g_def1} are derived algorithmically from equations \eqref{e:MSSschur1} and \eqref{e:MSSgrad} follwing the discrete adjoint approach as shown in appendix \ref{a:AdjMSS}. Note that the matrix on the left hand side of equation \eqref{e:MSSadjSchur1} is the same as the one for the tangent MSS KKT Schur complement \eqref{e:MSSschur1}. This is because the matrix $\uuline{\bA}$ is symmetric, and a symmetric matrix is self adjoint. 

The sensitivities of $\bar{J}$ can be computed from the adjoint solution $\hw(t)$ with the following expression

\begin{equation}
\frac{d\bar{J}}{ds} = \sum_{i=1}^K \int_{t_{i-1}}^{t_i} \pd{f}{s}\bigg|_{t}^* \hat{w}(t) \ dt
\label{e:MSSadjGrad}
\end{equation}

\noindent where

\begin{equation}
 \hat{w}(t) = \phi^{* \ t,t_i } P_{t_i} \hat{\bf w}_i  + \frac1T \left(\int_{t}^{t_{i}} \phi^{* \ t,\tau} \frac{\partial J}{\partial u}\bigg|_{\tau} \ d\tau \right) 
+ \frac1T \frac{\bar{J} - J_i}{f_i^T f_i} \phi^{* \ t,t_i} f_i, \qquad t_{i-1} \le t \le t_i
\label{e:MSSadjSoln}
\end{equation}

\noindent or, in differential equation form

\begin{gather}
-\dd{\hw}{t} = \pd{f}{u}\bigg|^*_t \hw + \frac1T \frac{\partial J}{\partial u}\bigg|_{t}, \qquad t_{i-1} \le t \le t_i, \nonumber\\
w(t_i) = P_{t_i} \hat{\bf w}_i  + \frac1T \frac{\bar{J} - J_i}{f_i^T f_i} f_i,
\label{e:MSSadjSolnOde}
\end{gather}

\noindent This can derived by using equation \eqref{e:b_def}, as shown in appendix \ref{a:AdjMSS}. 

Interestingly, the adjoint MSS KKT Schur complement \eqref{e:MSSadjSchur1} is a solution of the following minimization problem

\begin{equation}
\min_{\uhbw} \|\uuline{\bB}^T \uhbw + \ubg\|^2_2 + \epsilon \| \uhbw \|^2_2
\label{e:MSSadjTikhonov}
\end{equation}

\noindent This is called a ridge regression, or Tikhonov regularization of 

\[
\uuline{\bB}^T \uhbw = - \ubg 
\]

\noindent or, equivalently

\begin{equation}
\left(\begin{array}{cccc}
\Phi_1^T &  & & \\
 -I & \Phi_2^T &  &  \\
 & -I & \ddots &  \\
 & & \ddots & \Phi_K^T \\
 & & & -I 
\end{array}\right) \left(\begin{array}{c}
\hbw_1 \\
\hbw_2 \\
\vdots \\
\hbw_K
\end{array}\right) = -\left(\begin{array}{c}
\bg_1 \\
\bg_2 \\
\vdots \\
\bg_K \\
 0
\end{array}\right)
\label{e:OverConsAdj}
\end{equation}

\noindent this can also be written as

\begin{equation*}
\left(\begin{array}{cccccc}
-I & \Phi_1^T &  & & & \\
 & -I & \Phi_2^T & & &  \\
 & & -I & \ddots & & \\
 & & & \ddots & \Phi_K^T & \\
 & & & & -I & \Phi_{K+1}^T
\end{array}\right) \left(\begin{array}{c}
0 \\
\hbw_1 \\
\hbw_2 \\
\vdots \\
\hbw_K \\
0
\end{array}\right) = -\left(\begin{array}{c}
0 \\
\bg_1 \\
\bg_2 \\
\vdots \\
\bg_K \\
 0
\end{array}\right)
\end{equation*}

\noindent This shows that equation \eqref{e:OverConsAdj} can be interpreted as solving the conventional adjoint equation \eqref{e:MSSadjSoln} in the time horizon $t_0 < t < t_K$, with the initial and terminal conditions 

\[
\hw(t_0) = 0, \quad \hw(t_K) = 0
\]

As $\uuline{\bB}^T$ has more rows than columns, equation \eqref{e:OverConsAdj} is over-constrained and can be solved with some kind of regularization. The Tikhonov regularization is easy to analyze because unlike other popular regularizations it has an analytical solution for equation \eqref{e:MSSadjSchur1}, the adjoint MSS KKT Schur complement. 

\section{Implementation}
\label{s:mss_imp}

\subsection{Adjoint MSS Algorithm}
\label{ss:adjMSSalg}


Like transcription LSS implementations, the MSS implementation is comprised of a solver for the KKT Schur complement and a routine to compute gradients from the tangent or adjoint shadowing direction. For MSS, the KKT Schur complement matrix in equation \eqref{e:MSSschur1} or \eqref{e:MSSadjSchur1} is not formed explicitly. Instead, a multiple shooting algorithm is derived for equations \eqref{e:MSSschur1} or \eqref{e:MSSadjSchur1}. This section presents and discusses the multiple shooting algorithm for carrying out adjoint MSS. This algorithm was used to produce the results presented in this paper. The corresponding tangent MSS algorithm can be found in appendix \ref{a:tanMSSalg}. 

\begin{framed}

\noindent\textbf{Adjoint MSS Solver} \\
{\it Inputs:} Initial condition for the governing equations $u_0$, Spin-up time $t_0$, Specified time horizon and checkpoints $t_0,t_1,...,t_K$, Initial guess for the $Kn \times 1$ vector $\uhbw$, which contains the adjoint variables at checkpoints $1$ to $K$ (default value 0); \\
{\it Ouputs:} Sensitivities $d\bar{J}/ds$ \\
{\it Calls:} MATVEC algorithm that computes $\ubR = \uuline{\bA} \uhbw + \beta\uuline{\bB}\ubg$ where $\ubR$ is a $Kn \times 1$ residual vector. \\ 
\begin{enumerate}
\item Time integrate the governing equations \eqref{e:dyn_sys} to compute $u(t)$ for the specified time horizon. Save the objective function $\bar{J}$, it is needed for the right hand side of the linear system \eqref{e:MSSadjSchur1}. 
\item To form the right hand side of the linear system, $\uuline{\bB}\ubg$, use the MATVEC algorithm with $\beta = -1$ and $\ubw=0$. 
\item Use some iterative algorithm to solve equation \eqref{e:MSSschur1}. To compute the left hand side $\uuline{\bA}\uhbw$, use MATVEC with $\beta = 0$.  
\item Compute the sensitivity $d\bar{J}/ds$ using equation  \eqref{e:MSSadjGrad}.  
\end{enumerate}
\end{framed}

Next, a serial MATVEC algorithm is presented. Note that $t_i^-$ and $t_i^+$ refer to the time at checkpoint $i$ in time segments $i-1$ and $i$, respectively. For example, $\hw(t_i^+)$ is the adjoint solution at checkpoint $i$ in time segment $i$. 

\begin{framed}
\noindent\textbf{Serial Adjoint MATVEC Algorithm} \\
{\it Inputs:} $\uhbw$, a $Kn \times 1$ vector of the adjoint variables at checkpoints 1 to $K$; $\beta$, a scalar; \\
{\it Ouputs:} $\ubR$, a $Kn \times 1$ residual vector \\
MATVEC computes $\ubR = \uuline{\bA} \uhbw + \beta\uuline{\bB}\ubg$ \\
\begin{enumerate}
\item For all time segments, compute $\hw(t_{i-1}^+)$ by integrating
$\frac{d\hw}{dt} = -\left(\pd{f}{u}\right)^* \hw + \beta \frac1T\pd{J}{u},\;
t\in(t_{i-1},t_i)$ backwards in time with the terminal condition $\hw(t_i) = P_{t_i}\hbw_i$. 
\item Save $\hbv_{i-1} \equiv \hw(t_{i-1}^+) - \hbw_{i-1}$. If $i=1$, save $\hbv_0 \equiv \hw(t_0^+)$. 
\item For all time segments, compute $\hv'(t_i^-)$ by integrating
$\frac{d\hv'}{dt} = \pd{f}{u} \hv',\; t\in(t_{i-1},t_{i}]$
with the initial condition $\hv'(t_{i-1}) = \hbv_{i-1}$. 
\item For all time segments, compute $\bR_i = P_{t_i}\,\hv'(t_i^-) - \hbv_i + \epsilon \hbw_i$.  If $i=K$, $\hbv_K = - \hbw_K$. 
\end{enumerate}
\end{framed}

Some details of the above algorithms are discussed below. 

\subsection{Time Integration for MSS}
\label{ss:time_int}

Steps 1 and 3 of the MATVEC algorithm involve solving an adjoint equation for $\hw(t)$ and a tangent equation for $\hv'(t)$, respectively. The algorithm presented in this paper should work with any type of time integration scheme except multi-step schemes such as Adams-Bashforth and Adams-Moulton schemes. These scheme require multiple time steps for an initial condition, which the MATVEC algorithm presented in this paper does not support. Also, it is recommended that the scheme used for $\hw(t)$ is the discrete adjoint of the scheme used for $\hv'(t)$. 
Using discretely adjoint time integration schemes for solving the tangent and adjoint in MSS ensures that the matrix represented by the MATVEC algorithm is symmetric. 

Additionally, it is recommended that the time integration scheme used for computing $\hv'(t)$ is the same as that used for computing the solution to the governing equations $u(t)$. This makes the numerical solutions of $u(t)$ and $\hv'(t)$ discretely consistent. If the numerical solutions of $u(t)$, $\hv'(t)$, and $\hw(t)$ are all discretely consistent, then verification procedures similar to those used for transcription LSS \cite{Blonigan:2016:AIAA} can be used for MSS. If $u(t)$ and $\hv'(t)$ are not discretely consistent, but $\hv'(t)$ and $\hw(t)$ are computed by solvers with consistent and stable discretizations, then MSS should still compute accurate sensitivities, but the implementation would be more difficult to verify. 

It should be noted that a pair of time integration schemes for $\hv'(t)$ and $\hw(t)$ derived or computed using automatic differentiation will be discretely adjoint and discretely consistent with the scheme used to compute the governing equation solution $u(t)$. 

\subsection{Iterative Solver for MSS}

The 3rd step of the Adjoint MSS algorithm calls for some iterative solver for equation \eqref{e:MSSadjSchur1}. The implementation used for the results presented in this thesis uses MINRES, a Krylov subspace method for solving symmetric, sparse matrices \cite{Paige:1975:minres,Golub1996}. Another Krylov subspace method, conjugate gradient (CG), was tried on a few chaotic systems, but it was found to converge slightly slower than MINRES. Krylov subspace methods such as GMRES are also used by Sanchez and Net in their multiple shooting continuation algorithm \cite{Sanchez:2010:MS}. 

Krylov subspace methods are not the only viable methods to solve the adjoint MSS KKT Schur complement \eqref{e:MSSadjSchur1}. Other iterative solvers based on Multigrid-in-time could also be used, for example, Multigrid Reduction Methods (MGRIT) \cite{Friedhoff:2013:MG,Falgout:2015:mgrit,Falgout:2015:mgritCFD}. MGRIT uses a multiple shooting approach to solve PDEs, with coarse time step solutions on each time segment used to accelerate the convergence of a fine time step solution. Since MGRIT and MSS both incorporate multiple shooting, a promising area for future work is the study of MGRIT approaches for MSS. 

\subsection{Time-Parallel MSS}
\label{ss:mss_par}

The MATVEC algorithm presented in section \ref{ss:adjMSSalg} computes the adjoint and then the tangent on each time segment in sequence, but the computations of the adjoint and tangent on each segment can also be done in parallel. The MATVEC algorithm presented below is parallel-in-time and has each time segment assigned to one processor. 

\begin{framed}
\noindent\textbf{Time-Parallel Adjoint MATVEC Algorithm} \\
{\it Inputs:} $\uhbw$, a $Kn \times 1$ vector of the adjoint variables at checkpoints 1 to $K$; $\beta$, a scalar;  \\
{\it Ouputs:} $\ubR$, a $Kn \times 1$ residual vector \\
Parallel MATVEC computes $\ubR = \uuline{\bA} \uhbw + \beta\uuline{\bB}\ubg$ on $K$ processors. The $i$th time segment is assigned to processor $i$. Processor $i$ requires the  quantities $u(t)$ and $f(u(t);s)$ (or the ability to compute them) for $t\in(t_{i-1},t_i)$. \\
To start, processor $i$ has access to $\hbw_i$. \\
On processor $i$, do the following:
\begin{enumerate} 
\item Compute $\hw(t_{i-1}^+)$ by integrating
$\frac{d\hw}{dt} = -\left(\pd{f}{u}\right)^* \hw + \beta \frac1T\pd{J}{u},\;
t\in(t_{i-1},t_i)$ backwards in time with the terminal condition $\hw(t_i) = P_{t_i}\hbw_i$. Meanwhile, if $i<K$, send $\hbw_i$ to processor $i+1$. Also, if $i>1$, receive $\hbw_{i-1}$ from processor $i-1$. 
\item Compute and save $\hbv_{i-1} \equiv \hw(t_{i-1}^+) - \hbw_{i-1}$. If $i=1$, save $\hbv_0 \equiv \hw(t_0^+)$. 
\item Compute $\hv'(t_i^-)$ by integrating
$\frac{d\hv'}{dt} = \pd{f}{u} \hv',\; t\in(t_{i-1},t_{i}]$
with the initial condition $\hv'(t_{i-1}) = \hbv_{i-1}$. If $i=1$, use the initial condition $\hv'(t_0) = \hbv_0$. Meanwhile, if $i>1$, send $\hbv_{i-1}$ to processor $i-1$. Also, if $i<K$, receive $\hbv_i$ from processor $i+1$.
\item Compute $\bR_i = P_{t_i}\,\hv'(t_i^-) - \hbv_i + \epsilon \hbw_i$.  For $i=K$, $\hbv_K = - \hbw_K$. 
\end{enumerate}
\end{framed}

Due to the block tridiagonal structure of equation \eqref{e:MSSadjSchur1}, time-parallel MATVEC has a very straightforward processor communication pattern: processor $i$ only needs to communicate with processors $i-1$ and $i+1$. This means that time-parallel MATVEC will scale very well: the time required for time-parallel MATVEC should be roughly the same as that required to solve an adjoint and then a tangent equation over one time segment. For large dynamical systems or for processors with slow connections there might be some additional time needed to complete communication of $\hbw_i$ and $\hbv_i$. For a fixed time segment size, time-parallel MATVEC is much faster than serial MATVEC for large values of $K$ which requires time for $K$ adjoint solves, followed by $K$ tangent solves. For a fixed time horizon $T_1-T_0$, the run time of the time-parallel MATVEC will decrease if the number of processors increases with $K$. 

Note that scaling of a time-parallel MSS solver also depends on the linear solver used for step 3 of the Adjoint MSS solver algorithm. For instance, Krylov subspace solvers like MINRES, GMRES, and Conjugate Gradient require a reduce-all operation to compute dot products of the solution vector $\uhbw$ and residual $\ubR$ \cite{Paige:1975:minres,Golub1996,Saad:1986:GMRES}. Since reduce-all operations involve communication between all processors and one root processor, they have a negative impact on scaling of the linear solver and therefore MSS. 

The scaling of time-parallel MSS also depends on the spectral properties of the KKT system which are discussed in section \ref{s:mss_cond}. Assume there are one or more time segments per processor. For a fixed time-segment size and a non-zero value of $\epsilon$, the minimum and maximum eigenvalues of the KKT system will be independent of $K$. In this case parallel MSS should scale well if a Krylov solver is used because the condition number will stay the same, so the number of solver iterations should stay the same as well\footnote{This assumes the clustering of eigenvalues on the real line does not vary much with $K$. }. For a fixed time horizon size $T_1-T_0$ and a non-zero value of $\epsilon$, the maximum eigenvalue of the KKT system will decrease with $K$, while the minimum eigenvalue will be at least $\epsilon$. In this case, parallel MSS will scale well, since the time to run the MATVEC will decrease and a decrease in condition number means than the KKT system will likely need fewer solver iterations to converge. However, one must be careful of the impact of $\epsilon$ on the accuracy of sensitivities, which is discussed in section \ref{s:accuracy}. If $\epsilon=0$ parallel MSS may not scale well as the minimum eigenvalue will decrease as $K$ increases and the KKT system may require more solver iterations to converge. 

Finally, it should be mentioned that additional parallelization is possible for governing equations defined over space, like the Navier-Stokes equation. Space-parallel CFD solvers, such as FUN3D, have been in use for a number of years now. Using a space-parallel tangent and adjoint solver in steps 1 and 3 of time-parallel MATVEC would create a space-time-parallel MATVEC. In this case, ``processor $i$'' would actually refer to a group of processors corresponding to the tangent and adjoint solvers in time segment $i$. If MSS is to be extended to very large chaotic dynamical systems such as LES, a space-time-parallel MATVEC will be necessary to make the time-to-solution reasonable. Therefore, a space-time-parallel MATVEC should be investigated in future studies of MSS. 

\subsection{Memory requirements of MSS}
\label{ss:mss_mem}

The adjoint MSS solver algorithm can be implemented to make memory requirements more reasonable than those discussed for the transcription LSS implementation used in FUN3D \cite{Blonigan:2016:AIAA}. This implementation required storing the governing equation solution $u(t)$ and the linearization matrices, including $\partial f/\partial u$ and $\partial f/\partial s$, at all times in the time horizon studied. 

For MSS, major savings in memory usage could be realized by using checkpointing for the adjoint solver in step 3 of the MATVEC algorithm \cite{Wang:2009:Thesis}. Checkpointing allows one to time integrate the adjoint $\hw(t)$ backwards in time without saving the governing equation solution $u(t)$ and the linearization matrices. Further memory savings can be achieved by solving the tangent equation for $\hv'(t)$ in step 3 simultaneously with the governing equations. Such a scheme would only require storage space for $\hv'(t)$, $u(t)$, and the linearization matrices required for one time step. Both of these strategies allow for relatively low memory usage when computing the solution of the adjoint MSS KKT Schur complement equation \eqref{e:MSSadjSchur1}. 

Computing sensitivities using equation \eqref{e:MSSadjGrad} could also be made memory efficient by computing the integral 

\[
\int_{t_{i-1}}^{t_i} \pd{f}{s}\bigg|_{t}^* \hat{w}(t) \ dt
\]

\noindent on the fly as $\hw(t)$ is solved, instead of solving for $\hw(t)$, storing it, then computing the sensitivities of $\bar{J}$ to different input parameters $s$. 

MSS could be implemented by recomputing $u(t)$ on the fly while solving for $\hw(t)$ and $\hv'(t)$. This would only require storing $K$ primal solutions, one for each time segment. Transcription LSS requires $u$ at all $m$ time steps discretizing the time horizon $T_1-T_0$.  Therefore, MSS will require less memory than transcription LSS since $K<m$. 

Finally, the time-parallel MATVEC presented in section \ref{ss:mss_par} allows for using distributed memory as the data needed for each time segment can be stored on different CPU's. 

Overall, the MSS formulation offers many more opportunities to minimize storage requirements than transcription LSS. However, memory efficiency alone does not make an algorithm computationally efficient. The time required to complete an algorithm is even more important in many cases. 

\section{Examples of MSS}
\label{s:examples}

\subsection{Dowell's plate}
\label{sss:dowell}

\begin{figure}[htb!]
\centering
\includegraphics[width=0.5\textwidth]{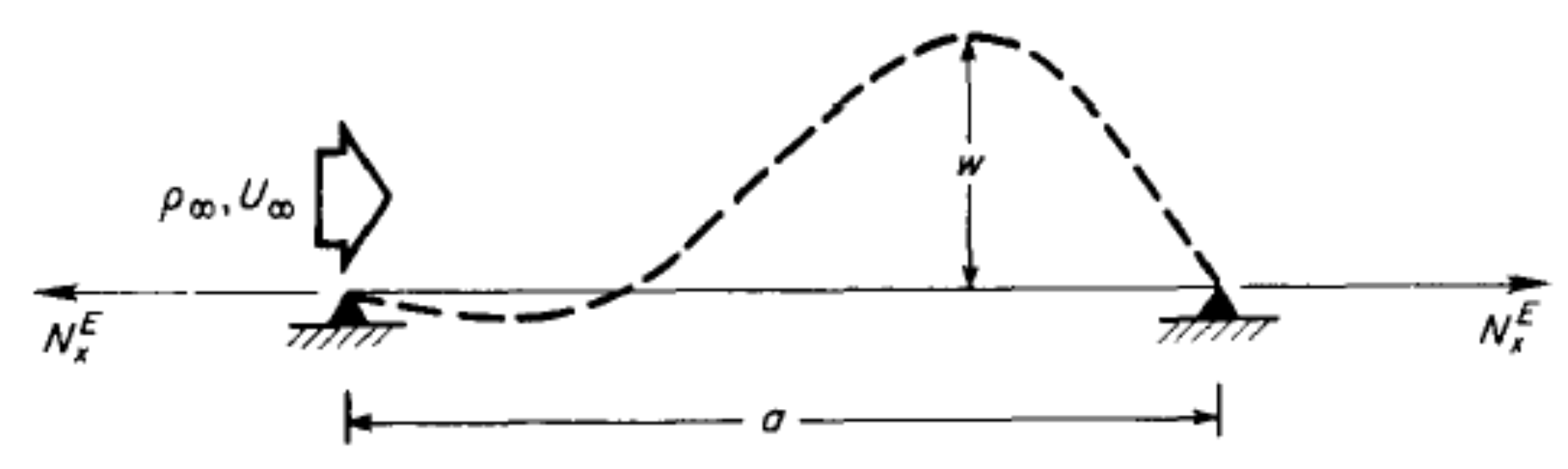}
\caption{Sketch of plate geometry. The magnitude of plate deflection, $w$, is exaggerated for clarity. Reprinted from \cite{Dowell:1982:chaos_plate}, Copyright 1982, with permission from Elsevier and the author. }
\label{f:dowell_plate}
\end{figure}

To study characteristics of the MSS algorithm including speed and accuracy, Dowell's plate model is studied. This model is a relatively simple aeroelastic test problem first explored by Dowell \cite{Dowell:1982:chaos_plate}. The set-up, shown in figure \ref{f:dowell_plate}, is a plate under a compressive in-plane loading $N_x^E$ in supersonic crossflow. For certain cases, Dowell observed chaotic flutter of this plate \cite{Dowell:1982:chaos_plate}. 

Using linear piston theory, a non-linear thin plate model, and assuming relatively high Mach numbers, the following PDE is obtained \cite{Dowell1975,Dowell:1966:AIAA_plate,Dowell:1967:AIAA_plate}:

\begin{equation}
D \pd{^4 w}{x^4} - (N_x + N_x^E) \pd{^2 w}{x^2} + m \pd{^2 w}{t^2} + \frac{\rho_{\infty} U_{\infty}^2}{M} \left[\pd{w}{x} + \frac{1}{U_{\infty}} \pd{w}{t} \right] = \Delta p
\label{e:dowell_pde}
\end{equation}

\noindent where the variables are defined in table \ref{t:dowell_vars}.

\begin{table}[htb!]
\centering
\begin{tabular}{|c|c|}
\hline Variable & Definition \\\hline\hline
$w$ & plate transverse deflection \\\hline
$x$ & streamwise spatial coordinate \\\hline
$t$ & time \\\hline
$a$ & plate length \\\hline
$h$ & plate thickness \\\hline
$m$ & mass per unit length of the plate \\\hline
$N_x^E$ & externally applied in-plane load (positive in tension) \\\hline
$E$ & modulus of elasticity \\\hline
$\nu$ & Poisson ratio of the plate material \\\hline
$D$ & plate bending stiffness \\\hline
$N_x$ & the tension created by the stretching of the plate due to bending \\\hline
$\rho_{\infty}$ & fluid mass density \\\hline
$U_{\infty}$ & flow velocity \\\hline
$M$ & flow mach number \\\hline
$\Delta p$ & static pressure difference across the plate \\\hline
\end{tabular}
\caption{Variable definitions for Dowell's buckled plate \cite{Dowell:1982:chaos_plate}. Note that $D=Eh^3/12(1-\nu^2)$ and $N_x = (Eh/2a) \int_0^a (\partial w/\partial x)^2 \ dx$. }
\label{t:dowell_vars}
\end{table}

A set of ordinary differential equations (ODE's) can be obtained from equation \eqref{e:dowell_pde} using Galerkin's method with the modal expansion

\[
w = \sum a_n(t) \sin(n\pi x/a),
\]

\noindent which is consistent with simply supported boundary conditions, $w = \partial^2 w/\partial x^2 = 0$ at $x=0$ and $x=a$ \cite{Dowell:1982:chaos_plate}. This results in the following non-dimensional equations

\begin{align}
A_n(n &\pi)^4 /2 + 6(1-\nu^2) \left[\sum_r A_r^2 (r\pi)^2/2 \right] A_n (n\pi)^2/2 + R_x A_n (n\pi)^2/2 + A_n^{''}/2 \nonumber\\
& + \lambda {\sum_m \left[ nm/(n^2-m^2) \right] \left[ 1-(-1)^{n+m} \right] A_m + (\mu/M\lambda)^{0.5} A_n^{'} } = P[1-(-1)^n]/(n\pi), \nonumber\\
& n = 1,2,...,\infty
\label{e:dowell_ode}
\end{align}

\noindent where $A_n \equiv a_n /h$, $\lambda \equiv \rho_{\infty} U_{\infty}^3 a^3 /MD$, $\mu \equiv \rho_{\infty}a/m$, $R_x \equiv N_x^E a^2/D$, $P \equiv \Delta p a^4/Dh$, $\tau \equiv t(D/ma^4)^{1/2}$, and a prime denotes $\partial (\ )/\partial\tau$. Also define $W \equiv w/h$ for later use. All results presented in this paper use four modes ($n=4$), which is said to be sufficient by Dowell \cite{Dowell:1982:chaos_plate}. 

The two parameters studied are the flow velocity parameter $\lambda$ and the load parameter $R_x$. As $\lambda$ is increased in the absence of external compression ($R_x=N_x=0$), the plate will flutter. On the other hand, as $R_x$ is increased the plate will buckle. Chaotic behavior is observed with the right combination of $\lambda$ and $R_x$, and occurs due to coupling in the buckling and flutter instabilities \cite{Dowell:1982:chaos_plate}. MSS is tested in this chaotic region of parameter space ($\lambda,R_x$). 

The objective function used for this study is the variance of the dimensionless transverse deflection $W(0.75a,t)$. Also, all calculations presented for Dowell's plate use the adjoint version of MSS. 

Time integration of the governing and tangent equations was done with a 3rd order explicit Runge-Kutta scheme based on the scheme in chapter 6 of \cite{Yang:2011:Thesis},

\begin{align}
u'_j &= u_j + \Delta t c_1 \dd{u}{t}\bigg|_{u=u_j,t=t_j} \nonumber\\
u''_j &= u'_j + \Delta t \left[ c_2 \dd{u}{t}\bigg|_{u=u'_j,t=t_j+r_1\Delta t} + d_2 \dd{u}{t}\bigg|_{u=u_j,t=t_j} \right] \label{e:RKo3}\\
u_{j+1} &= u''_j + \Delta t \left[ c_3 \dd{u}{t}\bigg|_{u=u''_j,t=t_j+r_2\Delta t} + d_3 \dd{u}{t}\bigg|_{u=u'_j,t=t_j+r_1\Delta t} \right] \nonumber
\end{align}

\noindent where subscript $j$ corresponds to the $j$th discrete time step, $\Delta t$ is the time step size and

\begin{gather*}
c_1 = \frac{1}{2}, \: c_2 = \frac{1}{3}, \: c_3 = 1, \: d_2 = -\frac{1}{6}, \: d_3 = -\frac{2}{3} \\
r_1 = c_1, \: r_2 = c_1 + c_2 + d_2.
\end{gather*}

The corresponding dual consistent adjoint time stepping scheme was used for adjoint solver for the reasons discussed in section \ref{ss:time_int}. The time step size $\Delta t$ was selected to ensure stability and accuracy of the scheme, resulting in fairly small time steps for Dowell's plate, which is modeled by stiff ODE's.

\begin{figure}
 \centering
\includegraphics[width=0.48\textwidth]{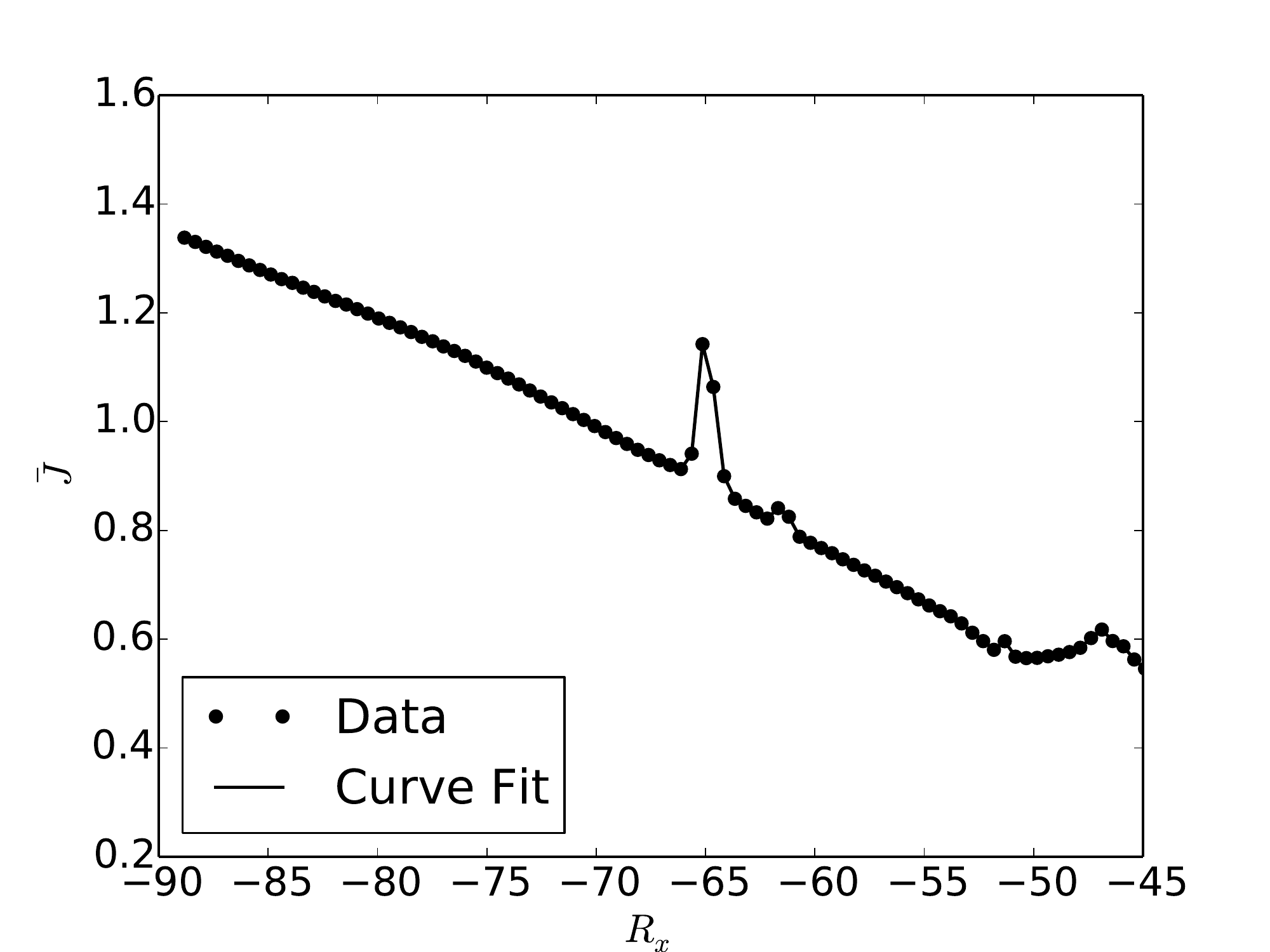}
\includegraphics[width=0.48\textwidth]{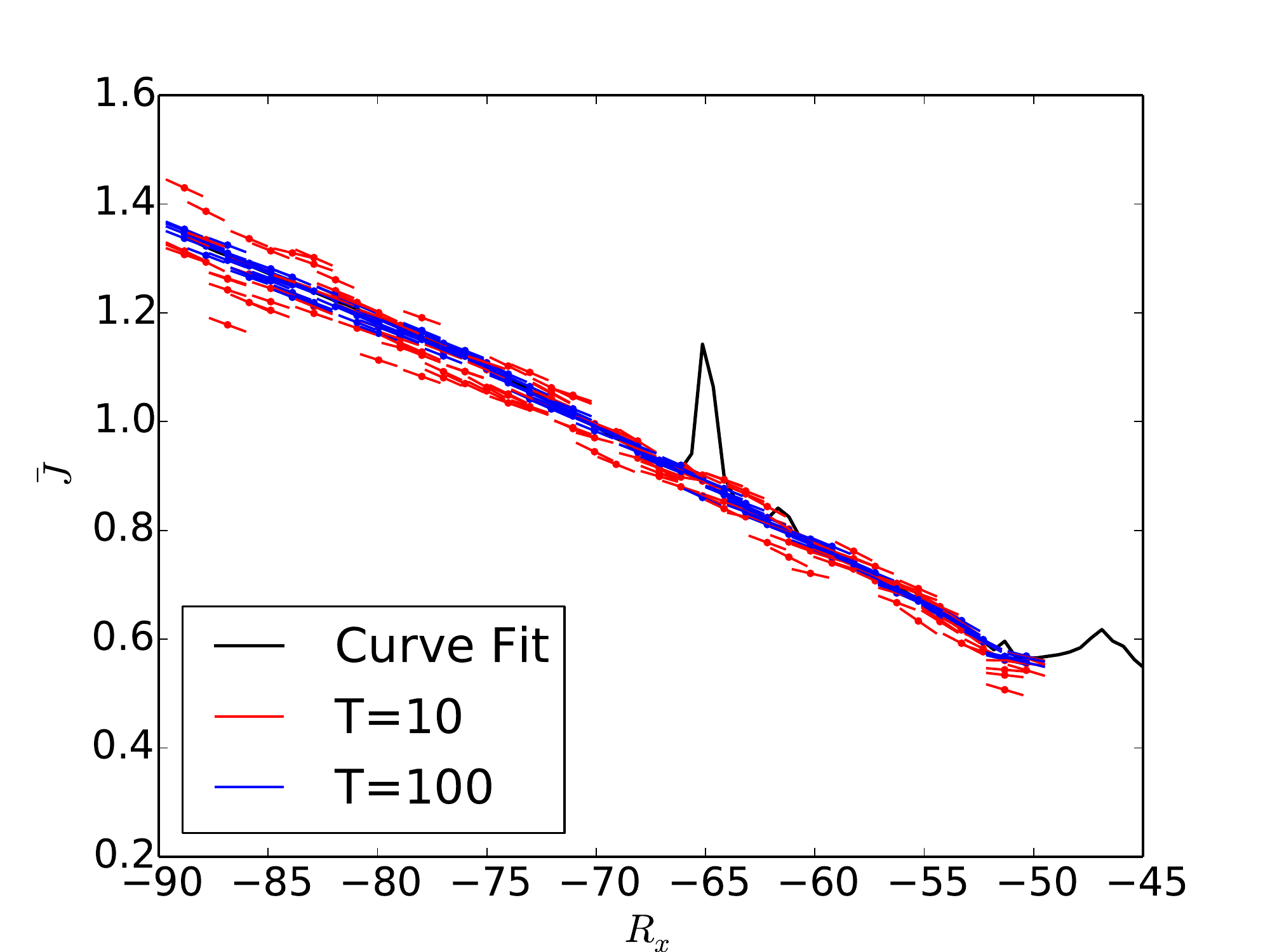}
\caption{LEFT: Objective function $\bar{J} = Var(W(0.75a,t))$ versus $R_x$ for $\lambda=150$. The black line shows a curve fit of $\bar{J}(s)$ computed with a 10,000 time unit time horizon, indicated by the black circles. The curve fit was conducted with monotone piecewise cubic interpolation \cite{Fritsch:1980:pchip}. RIGHT: $\bar{J}$ versus $R_x$. Sensitivities computed by MSS for two different time horizons are indicated by the slope of the lines at each point. Each simulation was run for 100 time units before averaging began from a distinct initial condition. Time segments of length $0.5$ were used for both time horizons. The curve fit from the top plot is also included. } 
\label{f:dowell_Rx_sweep}
\end{figure}

\begin{figure}
\centering
\begin{minipage}[t]{.48\textwidth}
  \centering
  \includegraphics[width=3.2in]{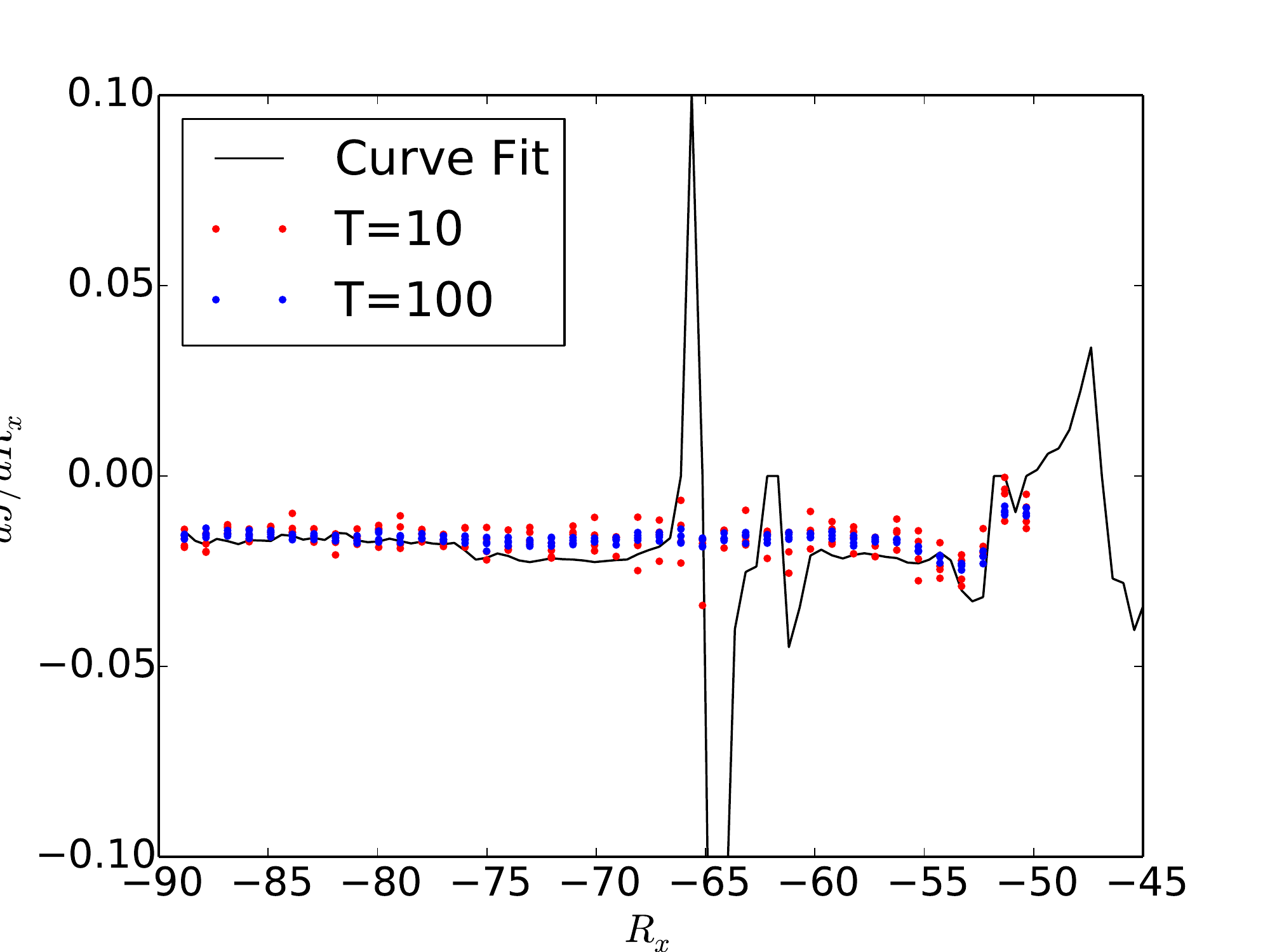}
  \caption{Sensitivities of $\bar{J} = Var(W(0.75a,t))$ with respect to $R_x$ versus $R_x$ for $\lambda=150$. The black line shows the derivative of the curve fit in figure \ref{f:dowell_Rx_sweep}. The MSS sensitivities are the same ones shown in figure \ref{f:dowell_Rx_sweep}. }
  \label{f:dowell_Rx_grads}
\end{minipage}
\hspace{0.1in}
\begin{minipage}[t]{.48\textwidth}
  \centering
  \includegraphics[width=3.2in]{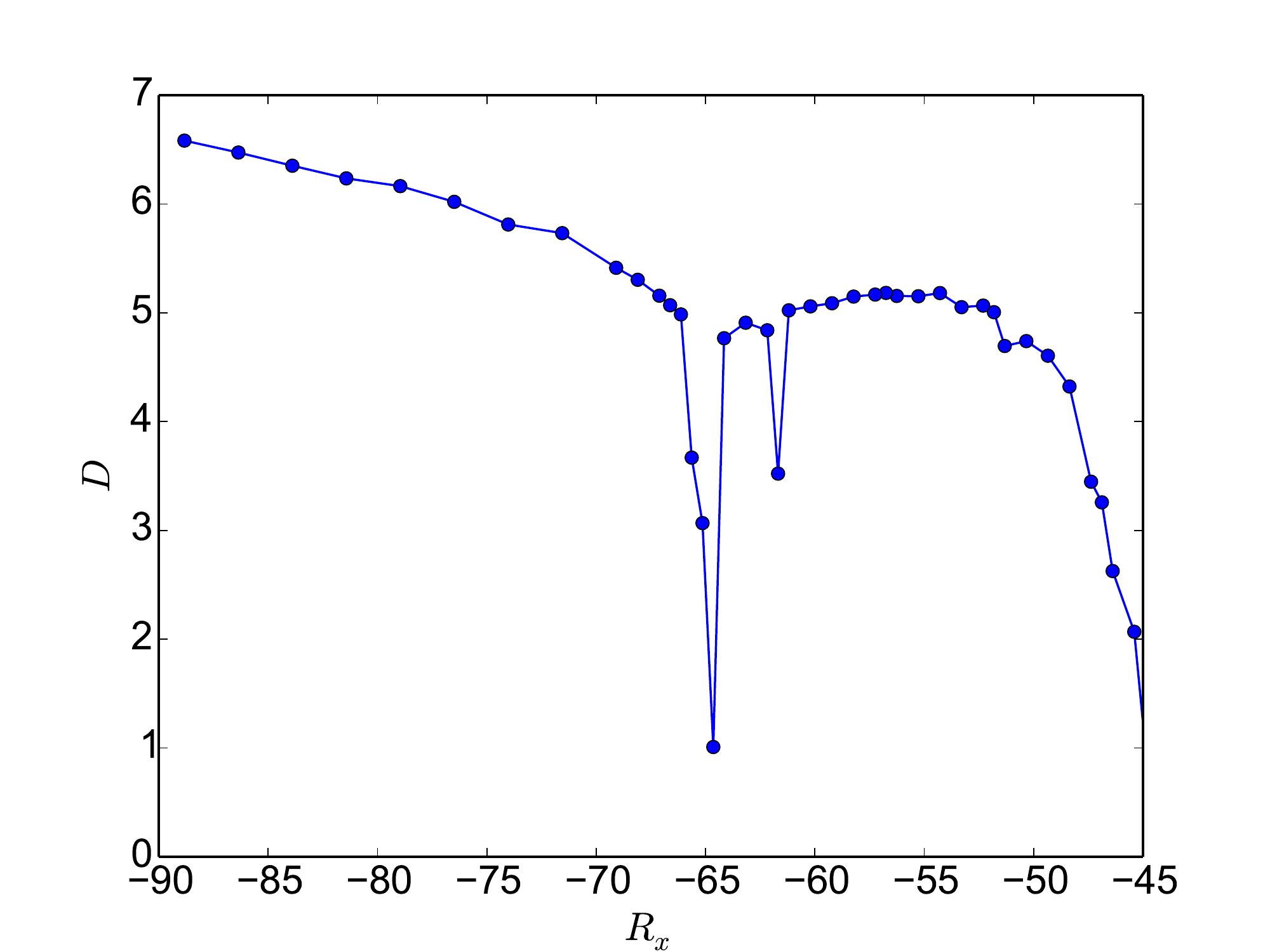}
  \caption{Attractor dimension $D$ versus $R_x$ for Dowell's plate (equation \eqref{e:dowell_ode}) with $\lambda=150$. The attractor dimension was computed with the Kaplan-Yorke conjecture \cite{Ruelle:1985:erg}. Lyapunov exponents were computed using the method of Benettin et al. \cite{Benettin:1980:Lyapunov}, with $s=1,000$ and $k=1,000$. }
  \label{f:Dowell_attr_dim}
\end{minipage}
\end{figure}

Figures \ref{f:dowell_Rx_sweep} and \ref{f:dowell_Rx_grads} show that MSS can compute accurate sensitivities for wide ranges of $R_x$. Figure \ref{f:dowell_Rx_grads} shows that MSS is very accurate for $R_x<-76.0$, fairly accurate for $-76.0<R_x<-68.0$ and $-60.0<R_x<-50.0$, and inaccurate for $-68.0<R_x<-60.0$. The regions where MSS is most inaccurate correspond to values of $R_x$ for which the attractor geometry is rapidly changing with respect to $R_x$, as shown in figure \ref{f:Dowell_attr_dim}. The rapid drops in attractor dimension around $R_x=-65.0$ and $R_x=-61.0$ are in the same region where the accuracy of MSS is degraded. This is consistent with the fact that sensitivity analysis is often ill-posed when attractor topology changes rapidly\footnote{Attractor topology changes such as bifurcations are inherently non-linear phenomena, so using using the tangent or adjoint equations to compute perturbations near them is a poor approximation, since the tangent and adjoint are linear. Also, there is no guarantee time-averaged objective functions will vary smoothly near topology changes \cite{Ruelle:1997:SRB}}. 

As MSS is accurate for a wide range of parameters for Dowell's plate, this test case is used to study the properties of MSS, including the conditioning of the KKT Schur complement matrix in equation \eqref{e:MSSadjSchur1}. 

\subsection{Kuramoto-Sivashinsky Equation}

\begin{figure}
 \centering
 \includegraphics[width=0.5\textwidth,trim = 5 45 5 70,clip]{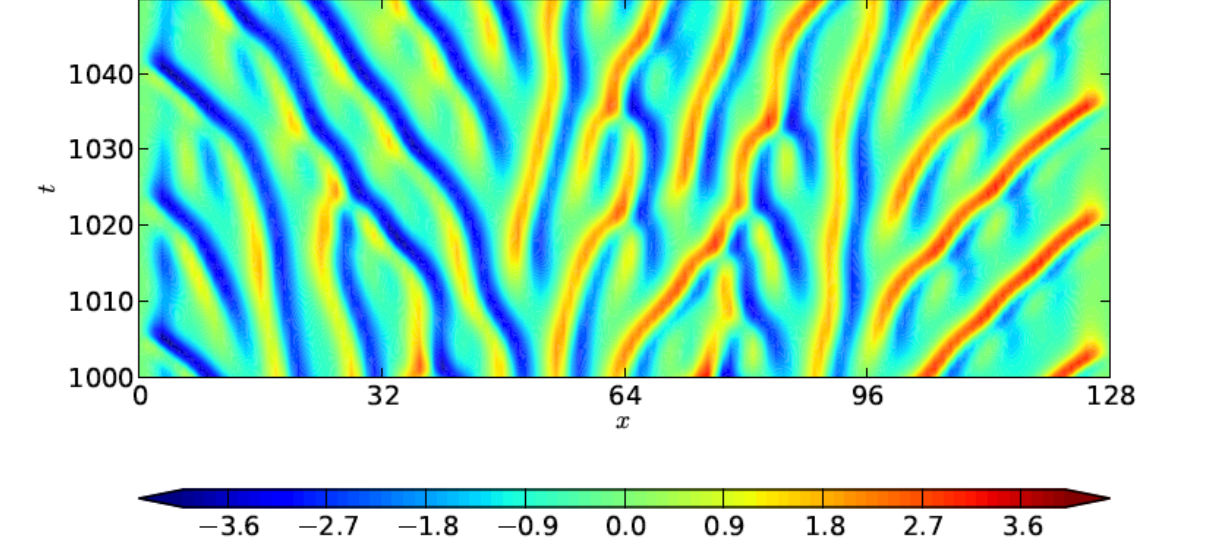}
 \caption{Typical solution of the K-S equation, equation \eqref{e:KS}, for $c=0.5$. The horizontal axis corresponds to the spatial dimension, the vertical axis to time. }
 \label{f:ks_primal}
\end{figure}

The second example is the Kuramoto-Sivashinsky (K-S) equation, a case previously studied with LSS \cite{Blonigan:2014:KS}. The K-S equation is a 4th order, chaotic PDE, that can be used to model a number of physical phenomena \cite{Hyman:1986:KS}. Kuramoto derived the equation for angular-phase turbulence for a system of reaction-diffusion equations modeling the Belouzov-Zabotinskii reaction in three spatial dimensions \cite{Kuramoto:1976:reaction,Kuramoto:1978:reaction}. Sivashinsky also derived the equation to model the evolution of instabilities in a distributed plane flame front \cite{Sivashinsky:1977:flames1,Sivashinsky:1977:flames2}. In addition the K-S equation has also been shown to be a model of Poiseuille flow of a film layer on an inclined plane \cite{Sivashinsky:1980:film}. Numerical studies typically use the 1D version of the K-S equation:

\begin{equation} 
\frac{\partial u}{\partial t} = 
-(u + c) \frac{\partial u}{\partial x}
- \frac{\partial^2 u}{\partial x^2}
- \frac{\partial^4 u}{\partial x^4}
\label{e:KS}
\end{equation}

\noindent The $c$ term is added to make the system ergodic \cite{Blonigan:2014:KS}. The equation was solved on the domain $0 \le x \le 128$, with the boundary conditions: 
\[ u\Big|_{x=0,128} = \frac{\partial u}{\partial x}\bigg|_{x=0,128}
 = 0 \]
 
\noindent These boundary conditions make the K-S equation ergodic, unlike the periodic boundary conditions used in many studies. A typical solution of the K-S equation from a randomized initial condition is shown in figure \ref{f:ks_primal}. The initial condition is randomized as follows: since equation \eqref{e:KS} is discretized with a 2nd order central difference scheme, $u(x,0)$ at each node is set to a random number drawn from a uniform distribution between -0.5 and 0.5. Time stepping was conducted with the 3rd order Runge-Kutta scheme from equation \eqref{e:RKo3}. A time step size of 0.1 units was used for the results presented in this paper.  

The objective function $\bar{J}$ considered for the K-S equation in this paper is $u^2$ averaged over both space and time:

\begin{equation}
\bar{J} = \frac{1}{128(T_1-T_0)} \int_{T_0}^{T_1} \int_0^{128} u^2(x,t) \ dx \ dt
\label{e:KSobj}
\end{equation}

\begin{figure}
 \centering
\includegraphics[width=0.48\textwidth]{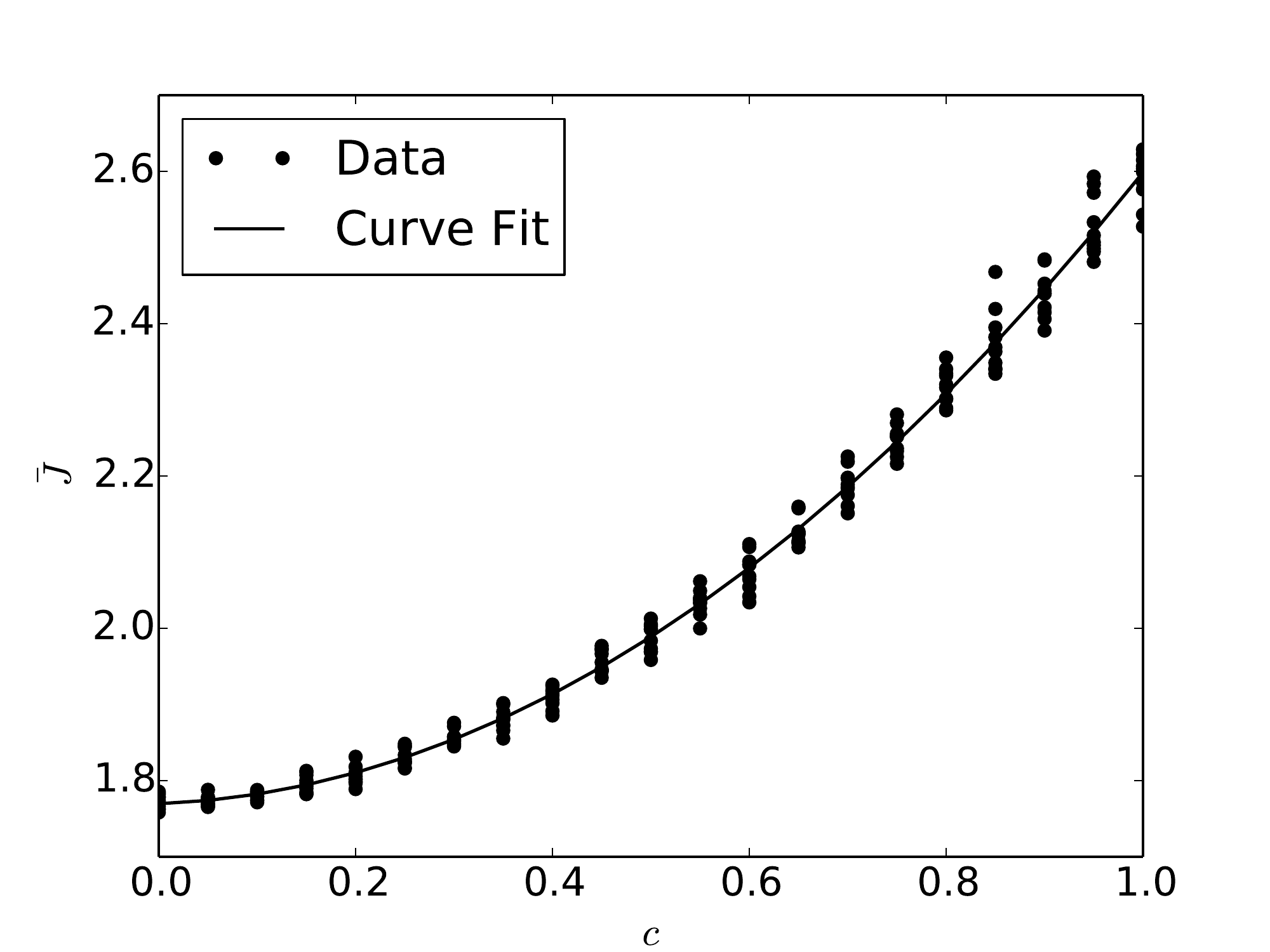}
\includegraphics[width=0.48\textwidth]{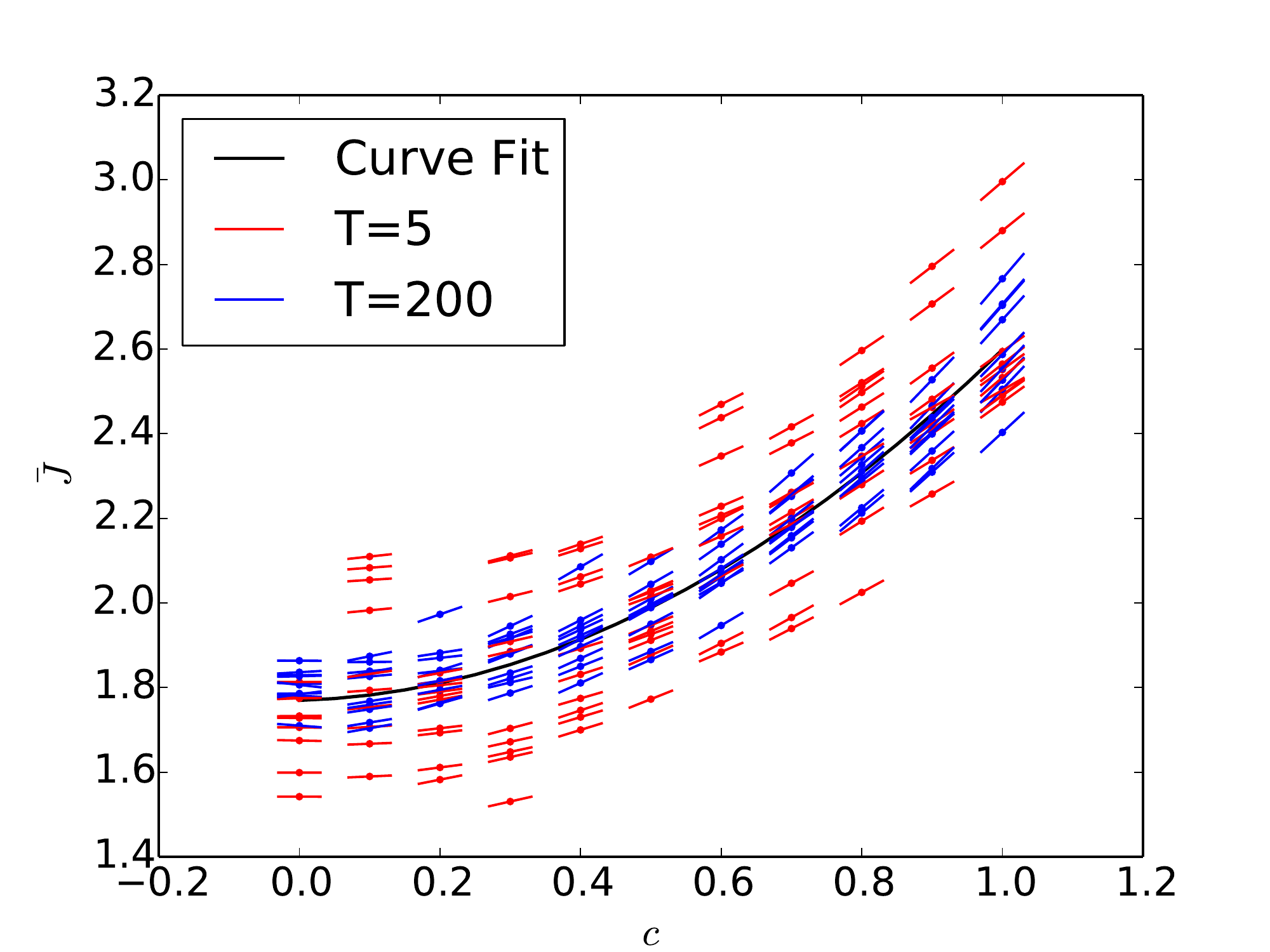}
\caption{LEFT: Objective function $\bar{J}$ (equation \eqref{e:KSobj}) versus $c$. The black line shows a curve fit of $\bar{J}(s)$ computed with a 10,000 time unit time horizon, indicated by the black circles. The curve fit was conducted with linear regression. RIGHT: Each simulation was run for 1000 time units before averaging began from a distinct initial condition. Sensitivities computed by MSS for two different time horizons are indicated by the slope of the lines at each point. Time segments of length $5.0$ were used for both time horizons. The curve fit from the left plot is also included. } 
\label{f:ks_sweep}
\end{figure}

Figure \ref{f:ks_sweep} shows sensitivities of computed by MSS. As for Dowell's plate, MSS computes accurate sensitivities for the K-S equation, just like LSS \cite{Blonigan:2014:KS}. 

\begin{figure}
  \centering
  \includegraphics[width=0.48\textwidth]{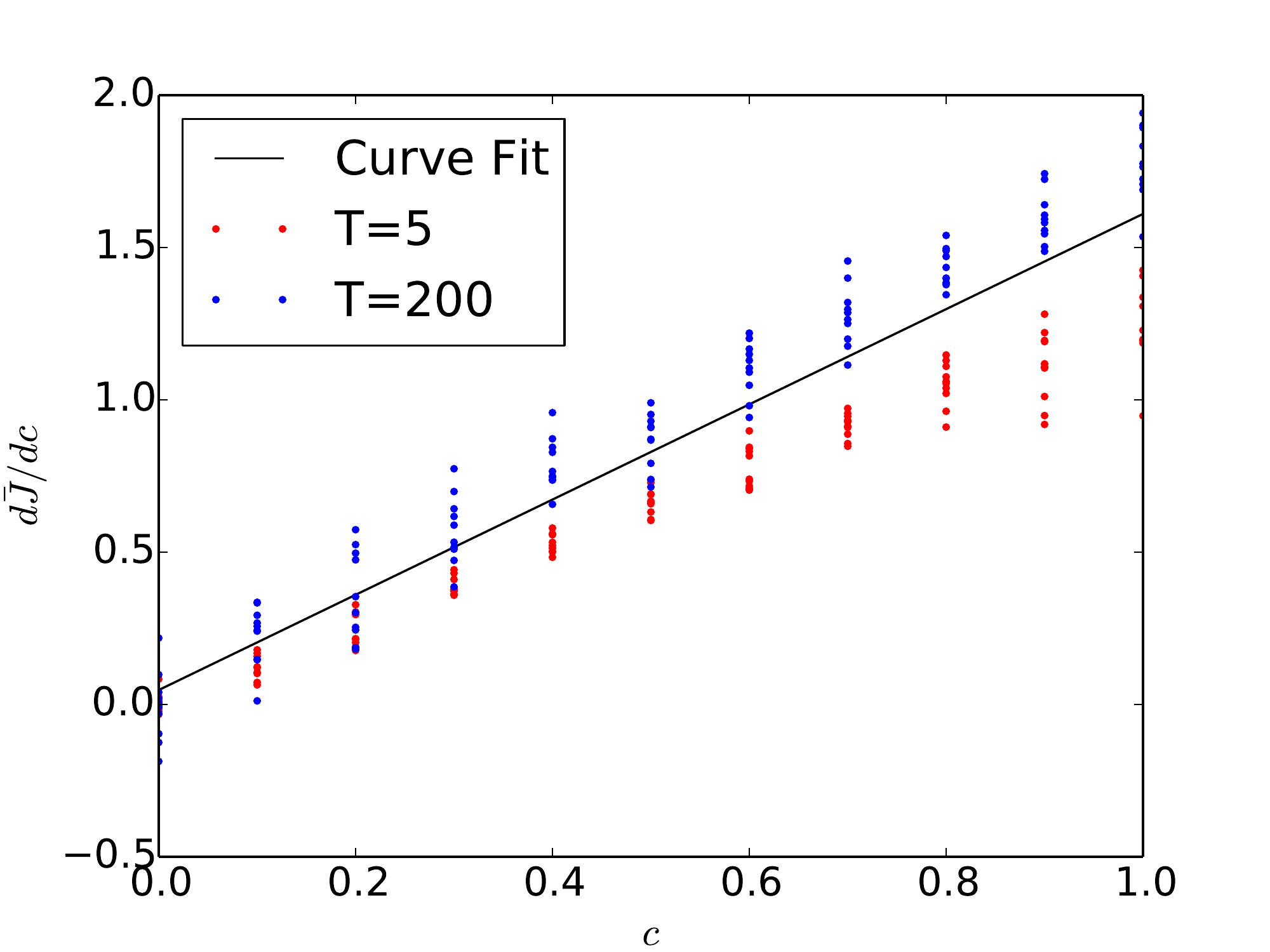}
  \caption{Sensitivities of $\bar{J}$ (equation \eqref{e:KSobj}) with respect to $c$ versus $c$. The black line shows the derivative of the curve fit in figure \ref{f:ks_sweep}. The MSS sensitivities are the same ones shown in figure \ref{f:ks_sweep}. }
  \label{f:ks_grads}
\end{figure}

\section{Rate of Convergence of MSS}
\label{s:mss_cond}

One of the main takeaways from the study of LSS for chaotic vortex shedding is that the transcription LSS implementation took a long time to compute sensitivities \cite{Blonigan:2016:AIAA}. This was mainly due to the large number of GMRES iterations (or search directions) required to converge the KKT Schur complement enough to obtain a sufficiently accurate sensitivity. To make LSS practical for large systems like those encountered in CFD simulations, convergence needs to be faster.


\begin{figure}
\centering
\begin{minipage}[t]{.48\textwidth}
  \centering
  \includegraphics[width=3.2in]{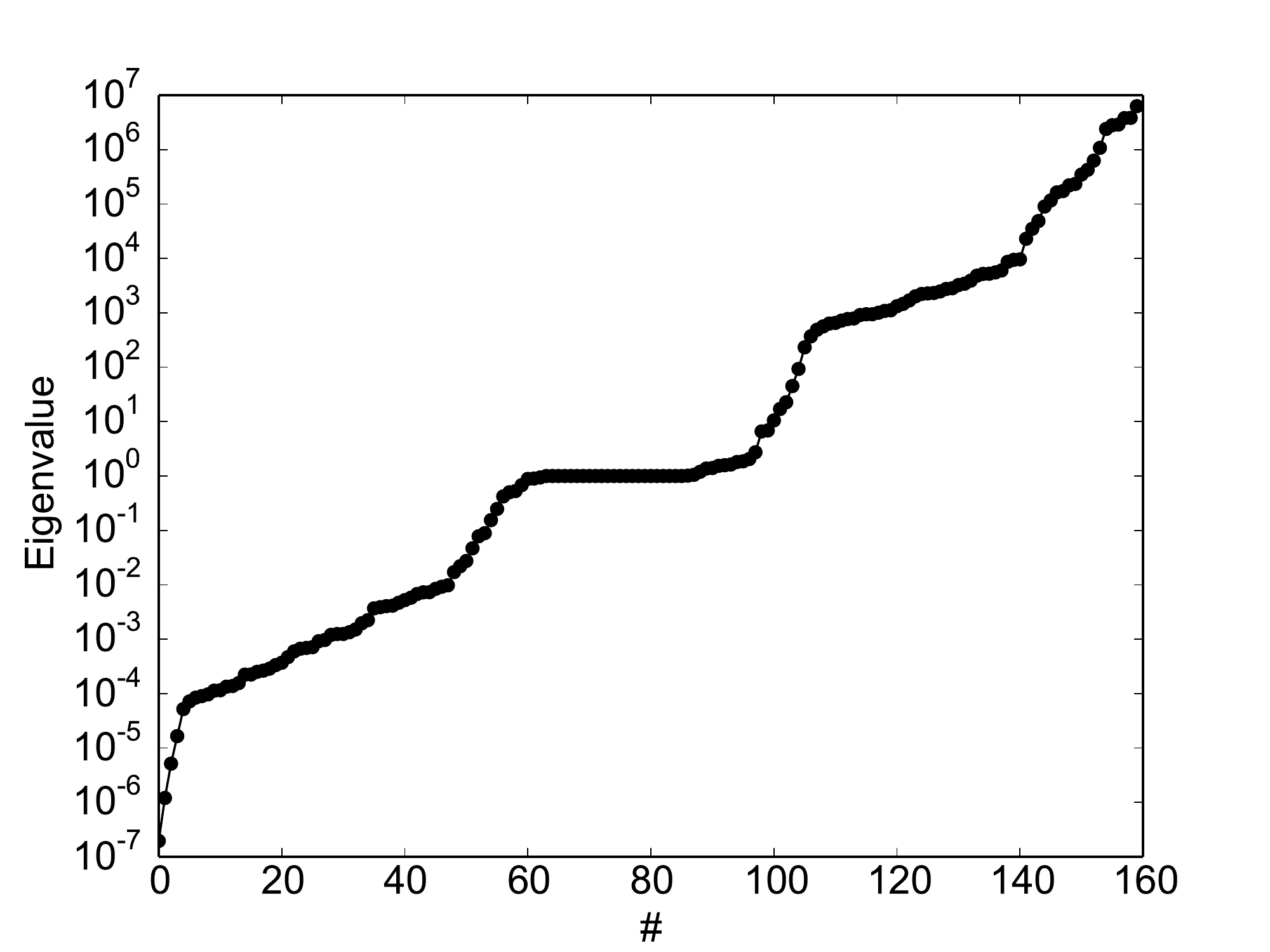}
  \caption{Typical example of an eigenvalue spectrum for the MSS KKT Schur complement matrix with $\epsilon=0$. This spectrum was computed for Dowell's plate, but spectra with similar features, most notably the wide range in magnitudes and a cluster of eigenvalues near $1$ have been computed for the K-S equation and Lorenz 63 system. This particular spectrum was computed for a simulation of Dowell's plate with the input parameters $(\lambda,R_x) = (150.0,-9\pi^2)$ and the initial condition $A_1 = 0.01$, $A_2 = A_3 = A_4 = 0.0$ and $A'_i = 0.0$ for $i=1,2,3,4$. MSS is applied with a spin-up time of $100$ units and a time horizon $T=10$ is used, with $K=20$. }
  \label{f:MSSspectrum}
\end{minipage}
\hspace{0.1in}
\begin{minipage}[t]{.48\textwidth}
  \centering
  \includegraphics[width=3.2in]{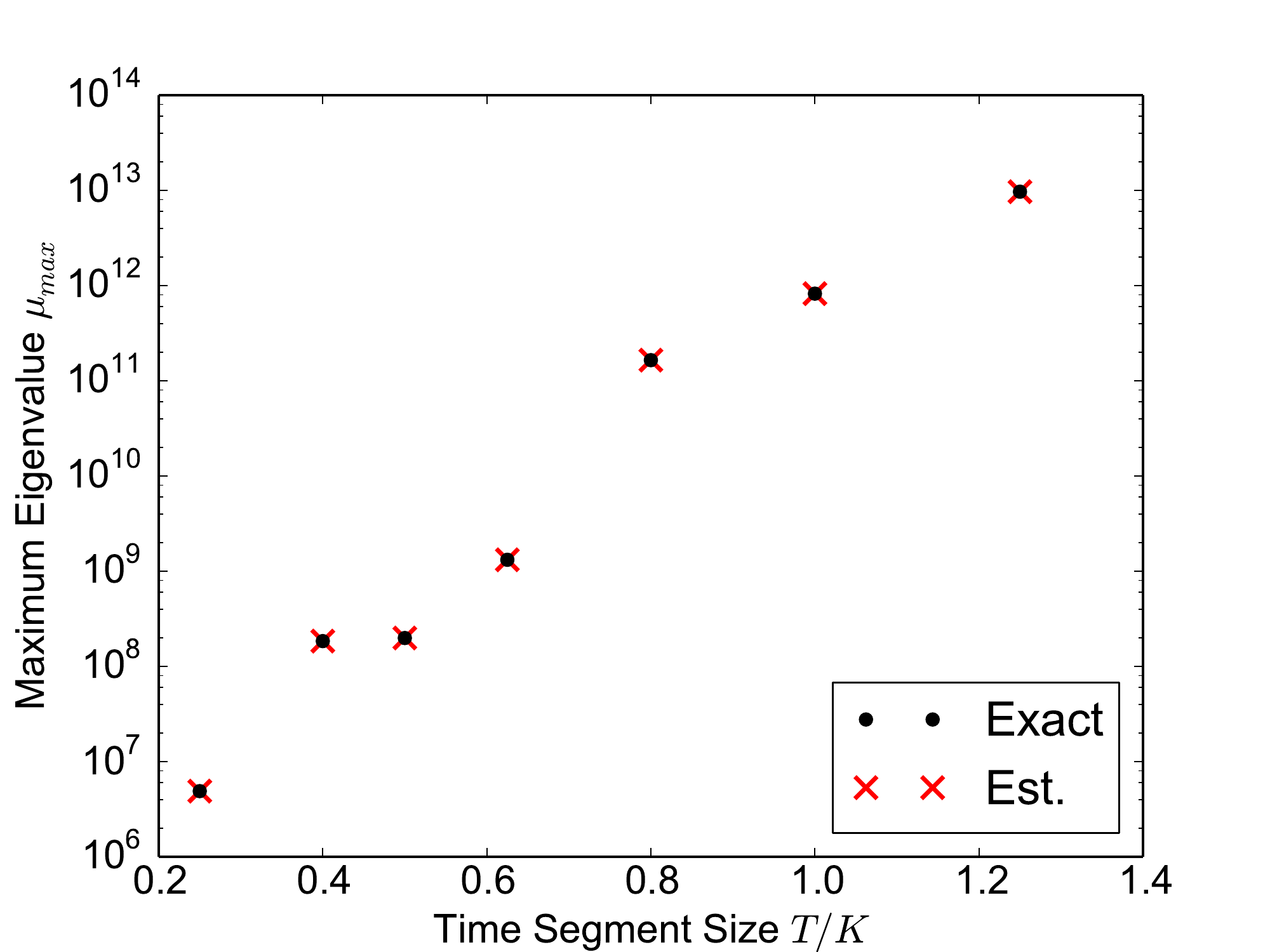}
  \caption{Maximum eigenvalue of the MSS KKT system versus time segment size $T/K$ for $T=100.0$ and $\epsilon = 0$. The estimated maximum eigenvalues were computed using equation \eqref{e:max_eig_lya}. These eigenvalues were computed for a simulation of Dowell's plate with the input parameters $(\lambda,R_x) = (150.0,-9\pi^2)$ and the initial condition $A_1 = 0.01$, $A_2 = A_3 = A_4 = 0.0$ and $A'_i = 0.0$ for $i=1,2,3,4$.  }
  \label{f:Lam_max_dT}
\end{minipage}
\end{figure}

The convergence rates of a wide range of iterative solvers depends on the condition number $\kappa$ of the matrix. For a symmetric positive definite matrix like the MSS Schur complement, the condition number $\kappa$ is defined as the ratio between the largest and smallest eigenvalues, $\mu_{max}$ and $\mu_{min}$. Typically, $\kappa$ is quite large for the KKT Schur complement matrices derived for both transcription LSS and MSS. Figure \ref{f:MSSspectrum} shows a typical spectrum for the MSS KKT Schur complement with $\epsilon=0$. This matrix has a condition number $\kappa\sim 10^{14}$. Large values of $\kappa$ are problematic, because iterative solvers typically converge faster for systems with smaller $\kappa$. This means that it is desirable to have $\mu_{max}$ and $\mu_{min}$ as close in value as possible. These eigenvalues depend on the chaotic dynamics of the system being analyzed, the number of time segments $K$, the time segment lengths, $\Delta T_i$, and the filtering parameter $\epsilon$. The details of the relationship of $\mu_{max}$ and $\mu_{min}$ to these properties and parameters is presented using some analysis of the MSS KKT Schur complement and some results for MSS applied to Dowell's plate. 

\subsection{The Largest Eigenvalue}
\label{ss:mss_mu_max}

First, the largest eigenvalue, $\mu_{max}$, is approximately (see \ref{a:MSS_max_ev})

\begin{equation}
\mu_{max} \approx 1 + \epsilon + \max_i e^{2\tilde{\Lambda}^{max}_i \Delta T_i}
\label{e:max_eig_lya}
\end{equation}

\noindent assuming $\max_i e^{\tilde{\Lambda}^{max}_i \Delta T_i} >>1$, where $\Delta T_i$ is the length of time segment $i$ and the quantity $\tilde{\Lambda}^{max}_i$ is the largest finite time Lyapunov exponent (FTLE) in time segment $i$. It defined as a finite time approximation of the largest Lyapunov exponent and 

\begin{equation}
\Lambda^{max} = \lim_{\Delta T_i \to \infty} \tilde{\Lambda}^{max}_i
\label{e:MSS_FTLya}
\end{equation}

The presence of $\tilde{\Lambda}^{max}_i$ in \eqref{e:max_eig_lya} shows that $\mu_{max}$ is linked to the dynamics of the governing equations being studied. It also explains why the magnitude of $\mu_{max}$ is typically large. For many chaotic dynamical systems, including chaotic fluid flows, it has been observed that FTLE's can vary widely on a strange attractor \cite{Sapsis:2013:locDim}. This means that in some time segments $\tilde{\Lambda}^{max}_i$ will be much larger than $\Lambda^{max}$, in others, it may even be negative. The former case can lead to large values of $\mu_{max}$ even for relatively small time segment lengths $\Delta T_i$. 


Figure \ref{f:Lam_max_dT} shows that the estimated value of $\mu_{max}$ according to equation \eqref{e:max_eig_lya} matches the true maximum eigenvalue very well. Also, it confirms that $\mu_{max}$ increases as the time segment size is increased. This growth in $\mu_{max}$ with time segment length $\Delta T_i$ is non-uniform because of the variation of the FTLE's on the regions of strange attractor encompassed by the MSS time horizon. Since the same solution $u(t)$ and time horizon is considered for all cases in figure \ref{f:Lam_max_dT}, different choices of time segments will compute different FTLE's. The time segment lengths are uniform, so $\Delta T_i = (t_K-t_0)/K$. Therefore, decreasing $K$ could decrease $\mu_{max}$ in some cases if the maximum FTLE $\tilde{\Lambda}^{max}_i$ decreases. However, these decreases are only temporary. As $K$ is decreased (and $\Delta T_i$ is increased) $\tilde{\Lambda}^{max}_i$ will approach $\Lambda^{max}$ and $\mu_{max}$ will grow exponentially with $\tilde{\Lambda}^{max}_i\Delta T_i$. 

Only uniform time segment lengths are considered in this thesis, but the variation of $\tilde{\Lambda}^{max}_i$ and its impact on $\mu_{max}$ may justify using non-uniform time segment lengths for MSS. Non-uniform time segment lengths could be used to control $\mu_{max}$ and perhaps even the condition number $\kappa$, depending on how non-uniform time segments impact the smallest eigenvalue $\mu_{min}$.

\subsection{The Smallest Eigenvalue}
\label{ss:min_eig}


\begin{figure}
\centering
\begin{minipage}[t]{.48\textwidth}
  \centering
  \includegraphics[width=3.2in]{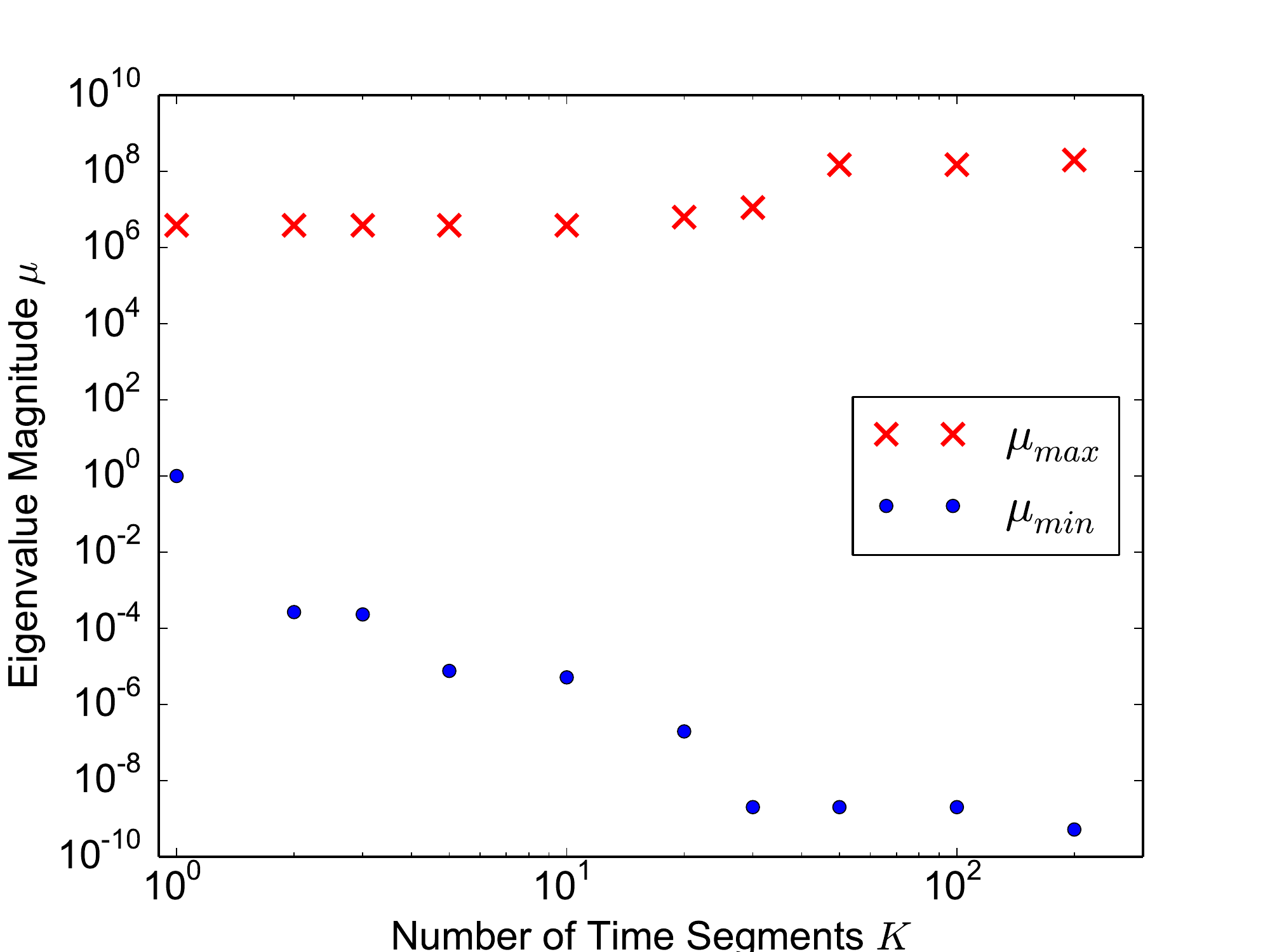}
  \caption{Maximum and Minimum eigenvalues of the MSS KKT system versus number of time segments $K$ for $T/K=0.5$ and $\epsilon = 0$. These eigenvalues were computed for a simulation of Dowell's plate with the input parameters $(\lambda,R_x) = (150.0,-9\pi^2)$ and the initial condition $A_1 = 0.01$, $A_2 = A_3 = A_4 = 0.0$ and $A'_i = 0.0$ for $i=1,2,3,4$.  }
  \label{f:Lam_min_K}
\end{minipage}
\hspace{0.1in}
\begin{minipage}[t]{.48\textwidth}
  \centering
  \includegraphics[width=3.2in]{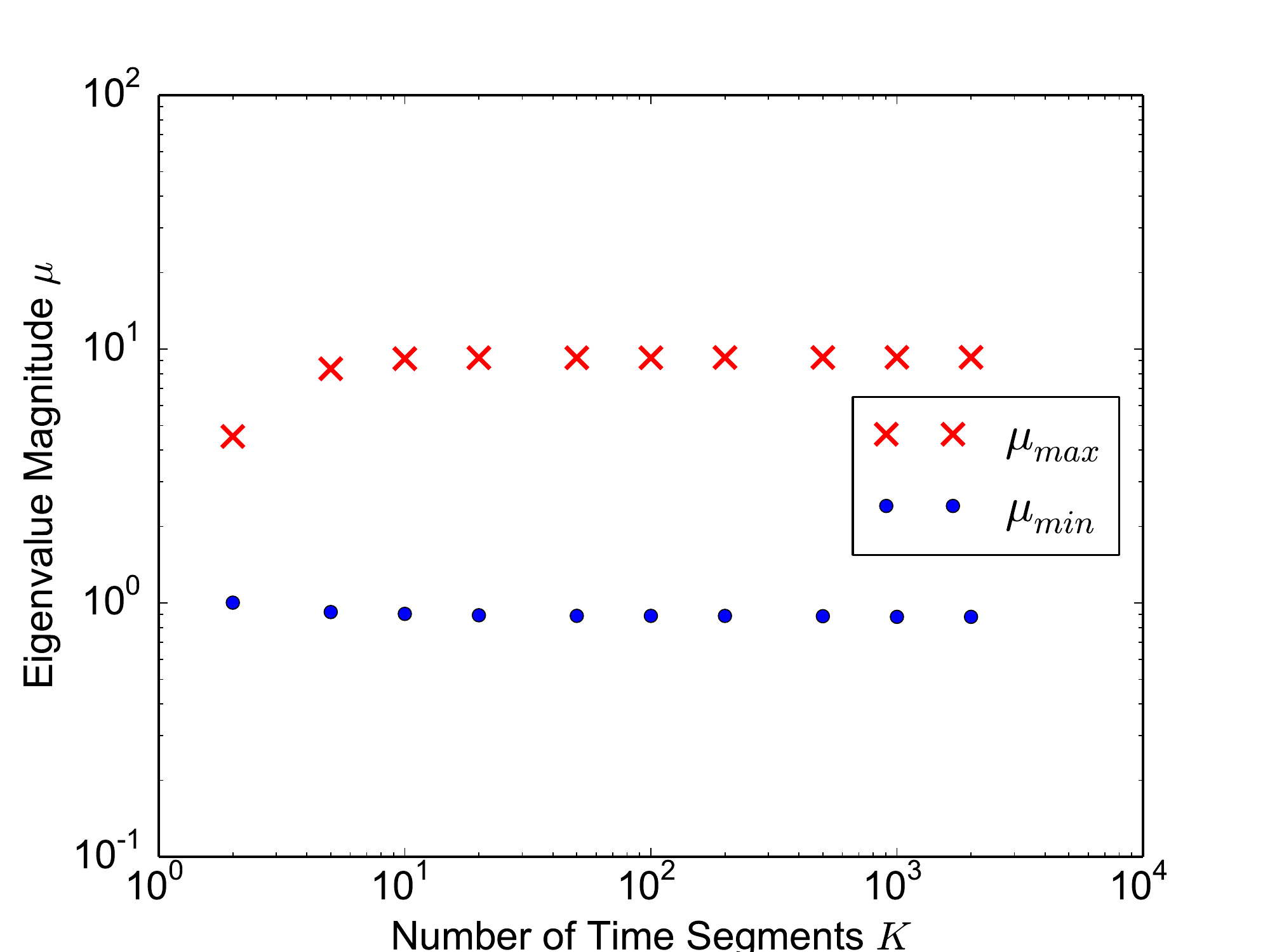}
  \caption{Maximum and Minimum eigenvalues of the MSS KKT system for the solenoid map \cite{Wang:2014:LSSthm} versus number of time segments $K$ for  $\epsilon = 0$. In this case, the number of time segments $K$ is equal to the number of map iterations $m$. This results in a KKT matrix identical to the one for MSS, since map functions map tangent states between discrete time planes, just like MSS does for differential equations. }
  \label{f:solenoid_lss_eigs}
\end{minipage}
\end{figure}

When LSS is conducted for a long enough time horizon, the KKT system for both MSS and transcription LSS can become nearly rank deficient, or $\mu_{min} \approx 0$.\footnote{This is the case for MSS when the filtering parameter $\epsilon$ is 0. } This issue was first observed for early applications of shadowing for noise control \cite{Farmer:1991:shadow}, but can arise for LSS as well, as shown by the very low values of $\mu_{min}$ in figure \ref{f:MSSspectrum}. Also the slow convergence of the KKT Schur complement system for the chaotic vortex shedding case studied by Blonigan et al. \cite{Blonigan:2016:AIAA} suggests the Schur complement is nearly rank deficient. 

The small magnitude of $\mu_{min}$ when $\epsilon=0$ is a symptom of applying LSS or any shadowing approach to a system with a quasi-hyperbolic attractor as opposed to a uniformly hyperbolic attractor. In fact, it has been shown that $\mu_{min}$ has a lower bound for uniformly hyperbolic attractors \cite{Wang:2013:LSS_PF}. As discussed in section \ref{s:past}, quasi-hyperbolic attractors include some points where some of the Lyapunov covariant vectors become parallel, called homoclinic tangencies \cite{Farmer:1991:shadow}. When a trajectory $u(t)$ passes very close to a homoclinic tangency, it is difficult to find corresponding shadow trajectories and therefore, shadowing directions $v(t)$ \cite{Farmer:1991:shadow}. 

To understand the difficulty of finding shadowing directions near homoclinic tangencies, consider the $i$th checkpoint for tangent MSS, where $v(t_i) = \bv_i$. If this checkpoint is on a hyperbolic region of the attractor, then $\bv_i$ and $v(t), t \in [t_i,t_{i+1}]$ can be expressed as a linear combination of Lyapunov covariant vectors, which act as a basis for all $n$ dimensions of phase space at the point $u(t)$. Define the minimum angle between any two Lyapunov covariant vectors at checkpoint $i$ as $\theta_i$. In the limit $\theta_i \to 0$, checkpoint $i$ coincides with a homoclinic tangency and two of the covariant vectors are parallel, so now $n-1$ of the $n$ covariant vectors are linearly independent. Therefore $\bv_i$ (and therefore $v(t), t \in [t_i,t_{i+1}]$) can only be expressed a linear combination of $n-1$ covariant vectors. This means there are only $n-1$ degrees of freedom to solve the checkpoint continuity constraint equation \eqref{e:MSSconstraint}. Since this is one of the equations that makes up the KKT system, the KKT system is also rank deficient. Therefore small $\theta_i$, or equivalently the close proximity of a checkpoint to a homoclinic tangency, implies a low condition number $\kappa$ for the entire system. 


This is problematic because longer time horizons $t_K - t_0$ will increase the likelihood of a checkpoint being close to a homoclinic tangency. This occurs because as a time horizon is increased in length, a solution $u(t)$ will pass near more homoclinic tangencies or pass closer by previously visited homoclinic tangencies. This trend was observed for a number of chaotic systems, including Dowell's plate, which is shown in figure \ref{f:Lam_min_K}. The minimum eigenvalue of the KKT system for Dowell's plate decreases by orders of magnitude as $K$ is increased. On the other hand, the minimum eigenvalue for the uniformly hyperbolic solenoid map shown in figure \ref{f:solenoid_lss_eigs}, stays approximately constant as $K$ is increased. This behavior is expected since hyperbolic attractors do not have homoclinic tangencies.

Finally, it should be noted that the decrease in the minimum eigenvalue due to homoclinic tangencies also explains the poor conditioning of the transcription LSS KKT system. This is because of the equivalence of MSS for $\epsilon=0$ and transcription LSS as $K\to\infty$ shown in section \ref{s:MSSform} and appendix \ref{a:MSS2tLSS}.





\subsection{Controlling the Condition Number with the Filtering Parameter}
\label{s:accuracy}



\begin{figure}
\centering
\includegraphics[width=0.48\textwidth]{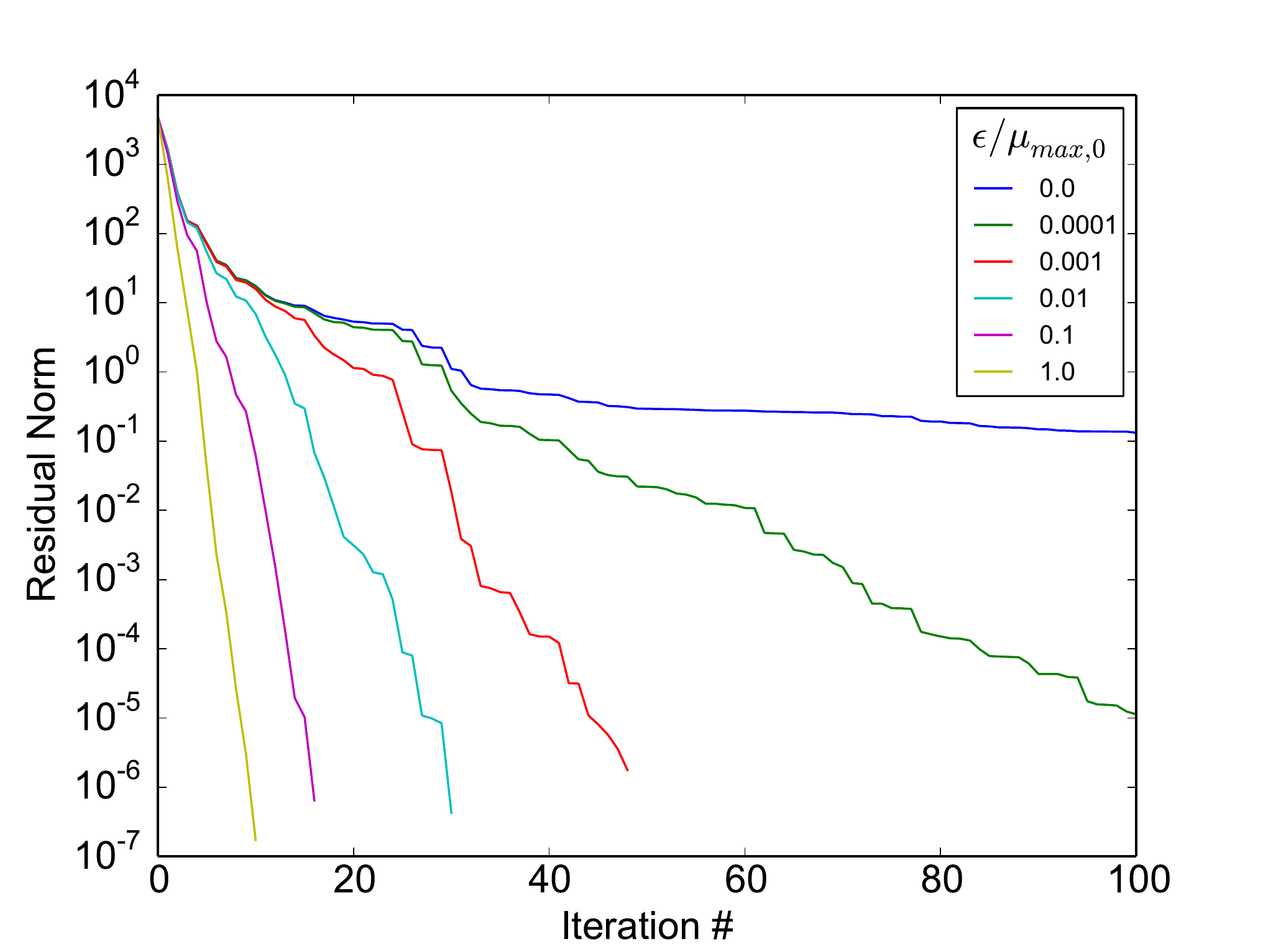}
\includegraphics[width=0.48\textwidth]{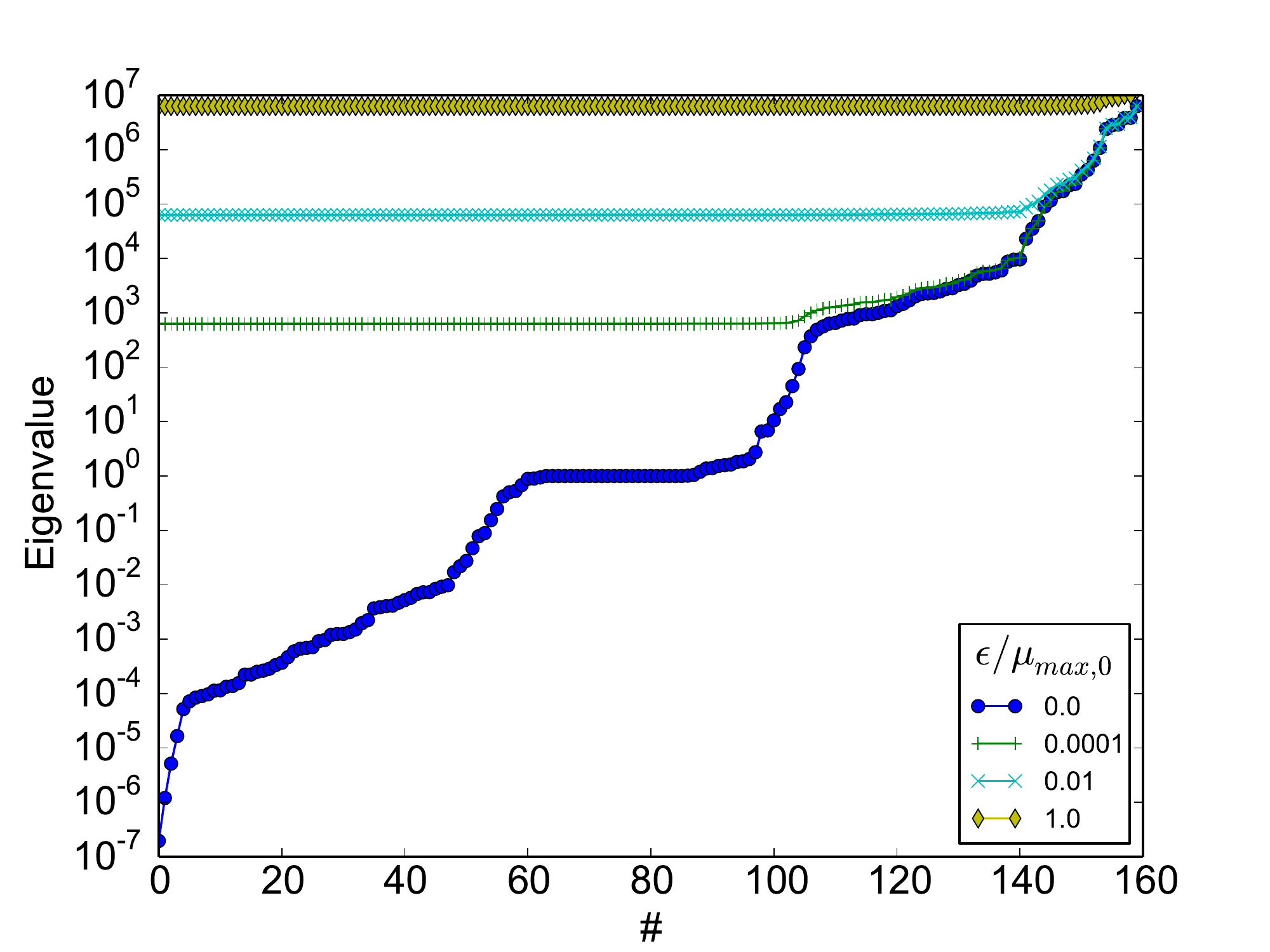}
\caption{LEFT: Convergence of the MSS residual norm $\| \uline{\bf r}\|_2$ for different values of $\epsilon/\mu_{max,0}$, where $\mu_{max,0}\approx 6.3 \times 10^6$ is the largest eigenvalue of $\uuline{\bA}$ when $\epsilon = 0$. The system is solved using MINRES \cite{Paige:1975:minres}. These residuals were computed for the simulation of Dowell's plate, with the same parameters, initial condition, and checkpoints used in figure \ref{f:MSSspectrum}. RIGHT: Correponding eigenvalue spectra for four values of $\epsilon/\mu_{max,0}$.  }
\label{f:MSScov_eps}
\end{figure}

The MSS filtering parameter, $\epsilon$, can be used to reduce the condition number of the MSS KKT system considerably. It does this by applying a constant shift $\epsilon$ to the entire eigenvalue spectrum. Define $\mu_{max,0}$ and $\mu_{min,0}$ as the maximum and minimum eigenvalues of the MSS KKT Schur complement with $\epsilon=0$, or $\uuline{B}\uuline{B}^T$. Then the condition number of the MSS KKT Schur complement with non-zero $\epsilon$ is:

\[
 \kappa = \frac{\mu_{max,0} + \epsilon}{\mu_{min,0} + \epsilon}
\]

\noindent Given that $\mu_{min}$ is typically much smaller than 1 for the MSS KKT Schur complement (see figure \ref{f:MSSspectrum}), a very small value of $\epsilon$ can improve the condition number $\kappa$ by orders of magnitude. This means that non-zero $\epsilon$ can reduce the number of iterations and therefore the amount of time required by an iterative solver to solve equation \eqref{e:MSSadjSchur1}. 


\begin{figure}
\centering
\begin{minipage}[t]{.48\textwidth}
  \centering
  \includegraphics[width=3.2in]{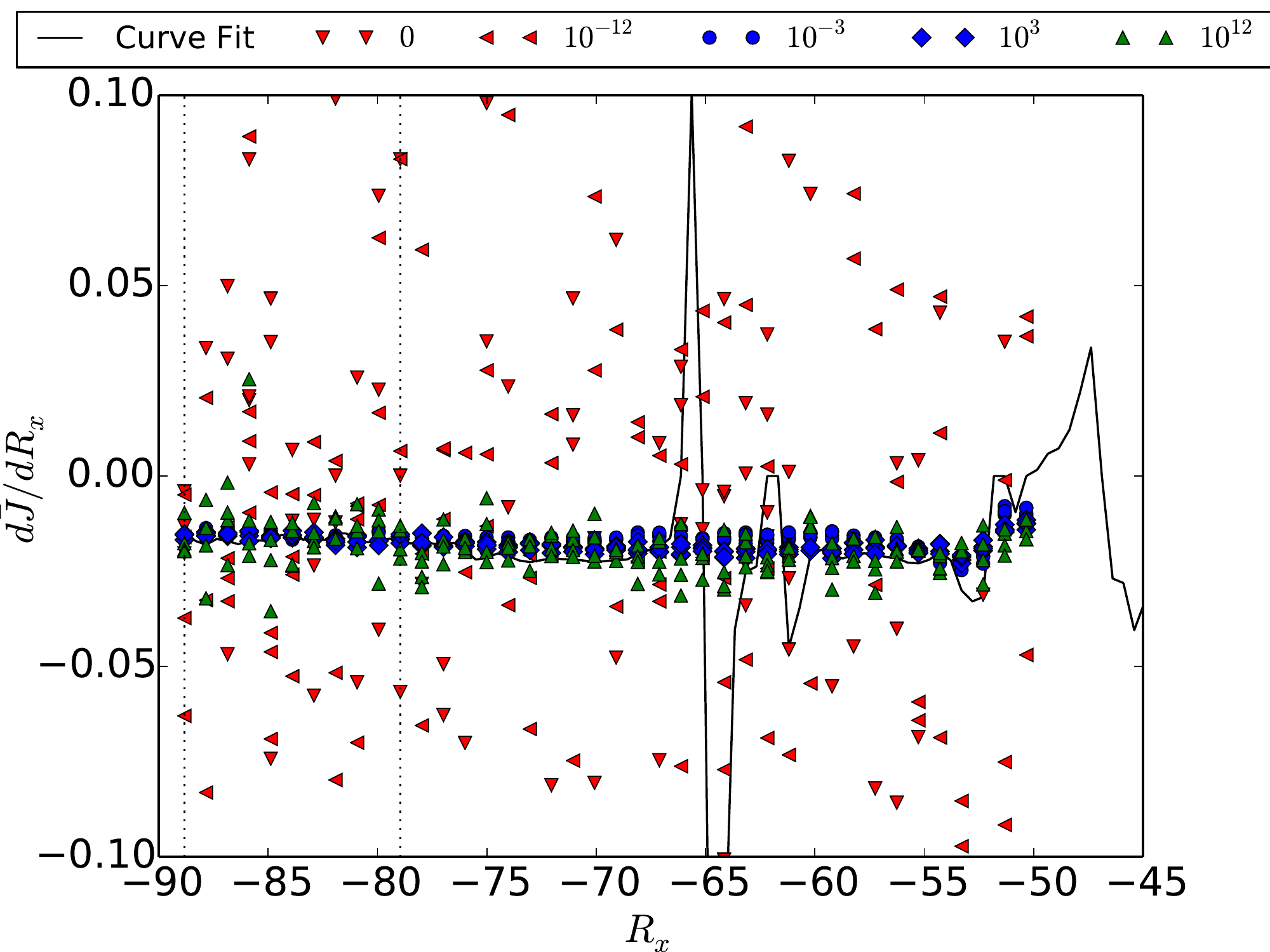}
  \caption{Sensitivities of $\bar{J} = Var(W(0.75a,t))$ with respect to $R_x$ versus $R_x$ for $\lambda=150$ and a range of different filtering parameter values $\epsilon$. The black line shows the derivative of the curve fit in figure \ref{f:dowell_Rx_sweep}. Dotted lines indicate the values of $R_x$ for which the sensitivity error is plotted in figure \ref{f:Dowell_grads_err}. All runs used a spin-up time of $T_0=100.0$ units, a time horizon of $T_1-T_0=100.0$ and $K=200$ segments. Sensitivities for 4 random initial conditions were computed for each values of $R_x$ and $\epsilon$. }
  \label{f:dowell_Rx_grads_eps}
\end{minipage}
\hspace{0.1in}
\begin{minipage}[t]{.48\textwidth}
  \centering
  \includegraphics[width=3.2in]{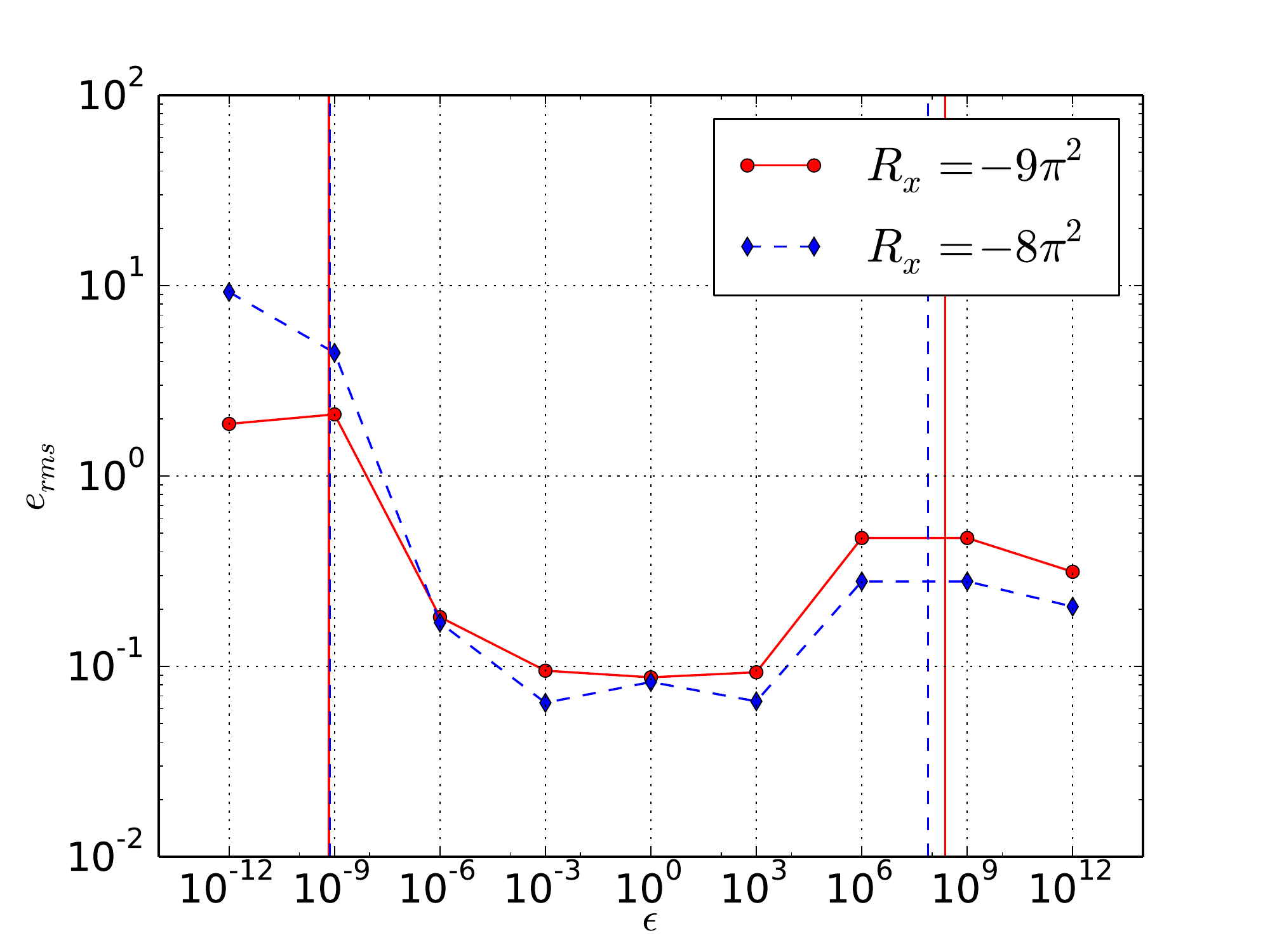}
  \caption{Root mean square error $e_{rms}$ of sensitivities shown in figure \ref{f:dowell_Rx_grads_eps} versus filtering parameter $\epsilon$ for two values of $R_x$. The average minimum and maximum eigenvalues for $\epsilon=0$, $\mu_{min,0}$ and $\mu_{max,0}$, are indicated by the vertical lines. The true sensitivity is estimated by the sensitivity computed by the derivative of the curve fit in figure \ref{f:dowell_Rx_sweep}.  }
  \label{f:Dowell_grads_err}
\end{minipage}
\end{figure}


This can be demonstrated on Dowell's plate. As the MSS system is shown to be nearly indefinite in figure \ref{f:MSSspectrum}, it is solved with the Krylov subspace solver MINRES \cite{Paige:1975:minres}. Figure \ref{f:MSScov_eps} shows that increasing $\epsilon$ results in faster convergence of the residual ${\bf r} =\uuline{\bA} \uhbw + \uuline{\bB}\ubg$. At the same time,figures \eqref{f:dowell_Rx_grads_eps} and \eqref{f:Dowell_grads_err} show that the choice of $\epsilon$ affects the sensitivities computed by MSS. These figures show that there is an optimal value of $\epsilon$ between the minimum and maximum eigenvalues $\mu_{min,0}$ and $\mu_{max,0}$. 

\begin{figure}
  \centering
  \includegraphics[width=0.48\textwidth]{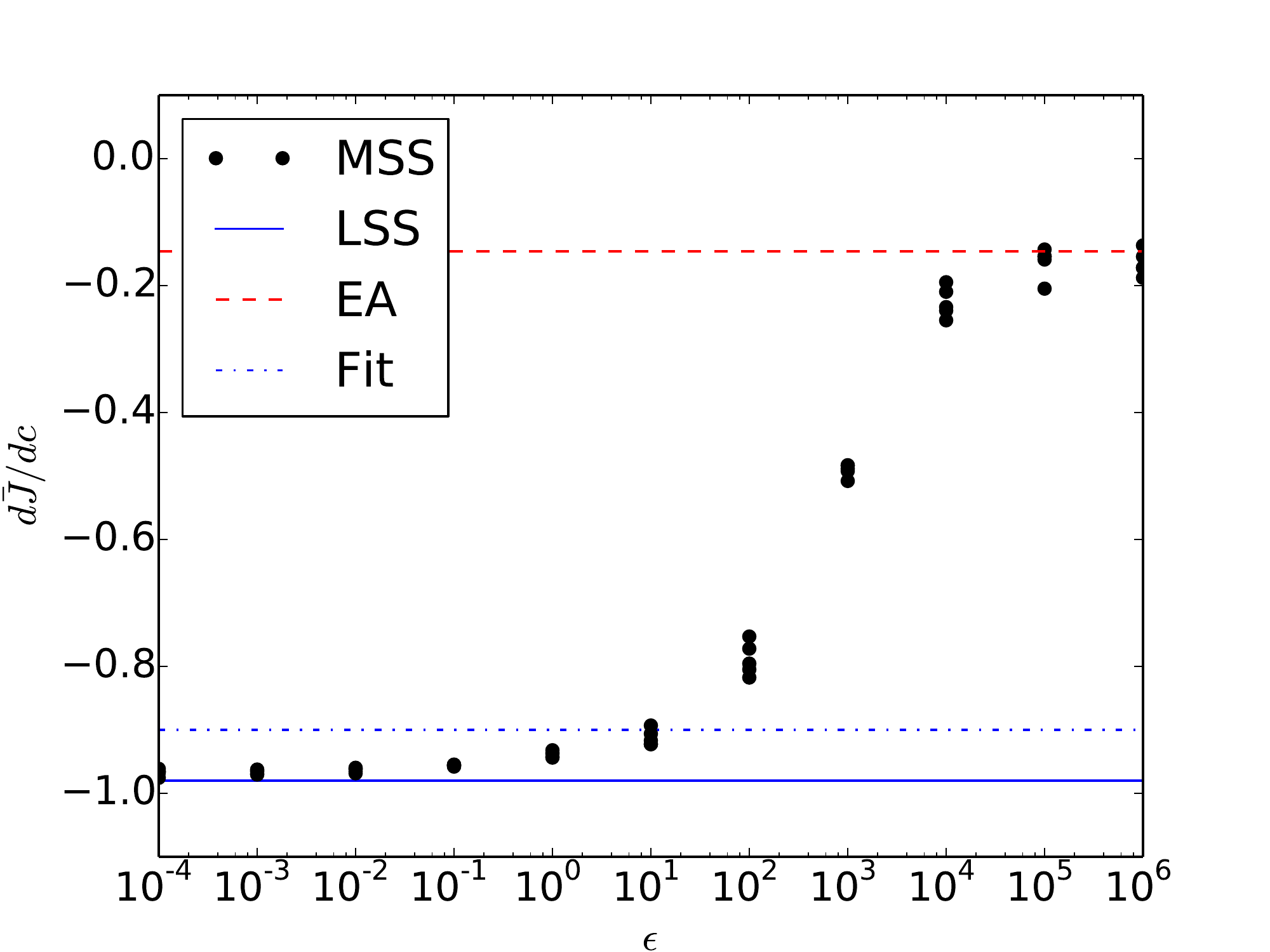}
  \caption{Sensitivity of $\bar{J} = \frac{1}{128(T_1-T_0)} \int_{T_0}^{T_1} \int_0^{128} u(x,t) \ dx \ dt$ with respect to $c$ for the K-S equation with $c=0.0$. MSS results are computed with a time horizon of $T = 102.4$ and 4 time segments. LSS and curve fit sensitivity obtained from \cite{Blonigan:2014:KS}. The ensemble adjoint (EA) sensitivity is computed with 44 samples of a $T=102.4$ time horizon.  }
 \label{f:KS_grads_eps}
\end{figure}

Figure \eqref{f:Dowell_grads_err} shows that the optimal value of $\epsilon$ is robust and a wide range of $\epsilon$ produces results with comparable accuracy for Dowell's plate. The errors when $\epsilon \approx \mu_{min,0}$ arise because of the poor conditioning of the minimization problem for Dowell's plate. These errors are not observed for the K-S equation, as shown in figure \ref{f:KS_grads_eps}. The errors at large values of $\epsilon$ observed in figures \ref{f:dowell_Rx_grads_eps} and \ref{f:Dowell_grads_err} arise because adjoint MSS becomes an ensemble adjoint approach as $\epsilon \to \infty$. For large values of $\epsilon$, consider equation \eqref{e:MSSadjSchur1}

\[
\uuline{\bA} \uhbw =  -\uuline{\bB}\ubg, \qquad \uuline{\bA} =\uuline{\bB}\uuline{\bB}^T + \epsilon \uuline{I}, 
\]

When $\epsilon$ becomes very large relative to the entries of $\uuline{\bB}\uuline{\bB}^T$ and $\uuline{\bB}\ubg$, equation \eqref{e:MSSadjSchur1} becomes 

\[
\epsilon \uhbw \approx 0
\]

This corresponds to solving the adjoint in each time segment with a terminal condition of 0, as in the ensemble adjoint method \cite{Eyink:2004:ensmbl}. The only difference is the presence of the projection operator $P_t$ in MSS, but this appears to have a small effect in practice, as seen in figure \ref{f:KS_grads_eps}. 

 The ensemble adjoint method requires a time segment around the size of the largest time scale of the system being analyzed \cite{Eyink:2004:ensmbl}. This time scale is proportional to the largest inverse Lyapunov exponent of the system. 

\[
t^{max} \sim \frac{1}{\Lambda^{|min|}}
\]

\noindent where $\Lambda^{|min|}$ is the Lyapunov exponent with the smallest non-zero magnitude. In the case of the modified K-S equation with $c=0.0$, this is roughly $1/0.004 = 250$ time units \cite{Blonigan:2014:KS}. Since the time segments used for the results in figure \ref{f:KS_grads_eps} are only 102.4 units, MSS suffers from the same bias as the ensemble adjoint. However, lower values of $\epsilon$ give sensitivities similar to those computed in a previous study \cite{Blonigan:2014:KS}.


\section{Conclusion} 
\label{s:conclusion}


 
Unlike conventional sensitivity analysis methods, LSS is able to compute useful sensitivities of long-time-averaged quantities in chaotic systems. Although other sensitivity analysis methods for chaotic systems have been developed in other communities, these are not viable approaches for high-fidelity, scale-resolving turbulence simulations like LES or DNS. The issues with these other methods include the slow convergence of the ensemble adjoint approach with the number of time horizons used, and the difficulties of extending Fokker-Planck approaches to chaotic systems with many DoFs. LSS avoids many of these issues, but the original implementation of LSS, transcription LSS, suffers from large computational costs. 

There are many ways the cost of LSS can be reduced by using multiple shooting rather than a transcription approach. The MSS implementation is definitely more memory efficient than transcription LSS, and can be made to converge faster with the right choice of time checkpoints and filtering parameter. Because of this, larger chaotic dynamical systems like scale-resolving fluid flow simulations should be analyzed with an MSS implementation. 

There is still work to be done on the MSS approach itself. This includes exploring other solver approaches such as MGRIT and a detailed investigation of preconditioners for MSS.

Overall, this work is imperative to realizing simulation based design for aerospace vehicles and other engineering devices which require high fidelity flow simulations for accurate analysis. Simulation based design with LSS would be very valuable, leading to gains in performance and providing a deeper understanding of the physics encountered by a wide range of vehicles and devices, from turbomachinery components to heavy-lift launch vehicles.

\appendix

\section{Computing the Maximum Lyapunov Exponent}
\label{a:lyapunov}


$\Lambda^{max}$ can be computed by the following approach, a simplification of the algorithm of  \cite{Benettin:1980:Lyapunov}:

\begin{framed}
\noindent\textbf{Compute $\Lambda^{max}$} \\
{\it Inputs:} Initial condition for the governing equations $u_0$, Spin-up time $t_0$, Specified time horizon and evenly spaced checkpoints $t_0,t_1,...,t_K$ with $t_i-t_{i-1} = \Delta t$; \\
{\it Ouputs:} Maximum Lyapunov exponent $\Lambda^{max}$ \\
\begin{enumerate}
\item Time integrate the governing equations \eqref{e:dyn_sys} to compute $u(t)$ for the specified time horizon.  
\item  Set  $i=1$ and set $\bv_{0}$ equal to some arbitrary unitary vector. 
\item Compute $v(t_i^-)$ by integrating
$\frac{dv}{dt} = \pd{f}{u} v \; t\in(t_{i-1},t_{i}]$
with the initial condition $v(t_{i-1}) = \bv_{i-1}$. 
\item Compute and save $R_i = \frac{\|v(t_i^-)\|_2}{\|v(t_{i-1})\|_2}$. 
\item Set $\bv_i = v(t_i^-)$. 
\item Set $i=i+1$ and repeat steps 3 to 5 until $i=K$
\item Solve 
\[
\Lambda^{max} \approx \frac{1}{t_K-t_0} \sum_{i=1}^K R_i
\]
This approximation becomes more accurate as $K \to \infty$ for any choice of $\Delta t$. \cite{Benettin:1980:Lyapunov}
\end{enumerate}

\end{framed}

\section{Relationship between MSS and Transcription LSS}
\label{a:MSS2tLSS}

For $\alpha=0$, the LSS minimization problem \eqref{e:opt_problem} becomes

\begin{equation}
 \min_{v} \frac{1}{2}\int_{T_0}^{T_1} \|v\|^2 \ dt, \quad s.t. \quad \frac{dv}{dt} = \frac{\partial f}{\partial u} v + \frac{\partial f}{\partial s} + \eta f \,, \quad T_0<t<T_1 ,
 \label{e:opt_a0}
\end{equation}

\noindent with the KKT equations

\begin{align}
\dd{w}{t} &= -\left(\frac{\partial f}{\partial u}\right)^* w - v, \qquad
w(T_0) = w(T_1) = 0 \label{e:KKTw_a0}\\
0 &= - \langle f, w \rangle \label{e:KKTeta_a0}\\
\frac{dv}{dt} &= \frac{\partial f}{\partial u} v + \frac{\partial f}{\partial s} + \eta f \label{e:KKTv_a0}
\end{align}

First, for uniformly spaced checkpoints $t_i$ and $\epsilon=0$, the MSS minimization statement multiplied by $\Delta t = t_{i+1}-t_i$ is

\begin{equation}
\lim_{K \to \infty} \frac{1}{2}\sum_{i=0}^{K-1} \left(\|v(t_i) \|^2 \Delta t \right) + \|v(t_K) \|^2 \Delta t = \frac12 \int_{t_0}^{t_K} \| v(t) \|^2 \ dt
\label{e:MSS_limit}
\end{equation}

\noindent For $t_0=T_0$ and $t_K=T_1$ this is equivalent to equation \eqref{e:opt_a0}. The limit in equation \eqref{e:MSS_limit} is true because a Riemann sum approaches the Riemann integral as smaller partitions $\Delta t$ are considered. This also ensures that the second term on the left of \eqref{e:MSS_limit} disappears as $\lim_{K \to \infty} \delta = 0$. 

The transcription LSS KKT equations are also consistent with the MSS constraint \eqref{e:MSSconstraint}. The KKT equations \eqref{e:KKTeta_a0} and \eqref{e:KKTv_a0} is equivalent to equations \eqref{e:MSStangent} and \eqref{e:MSSdilation_cond} when $\epsilon=0$ and continuity is enforced at checkpoints $t_i$. 

Equations \eqref{e:KKTw_a0} and \eqref{e:KKTeta_a0} imply equation \eqref{e:MSSdilation_cond} holds when $\alpha=0$. This is because $\langle f(u(t);s), v(t) \rangle = 0$ must also hold to ensure that $dw/dt$ and therefore $w(t)$ has no component parallel to $f(u(t);s)$ and equation \eqref{e:KKTeta_a0} is satisfied. 

Together, all of this shows that the MSS minimization problem is equivalent to the transcription LSS minimization problem as $K \to \infty$ if the checkpoints are uniformly spaced, $\epsilon = 0$ and $\alpha = 0$.

\section{MSS Gradient}
\label{a:MSSgrad}

The expression to compute the gradient from $v(t)$ and $\eta(t)$ is

\begin{equation}
\frac{d\bar{J}}{ds} = 
\frac1T \int_{t_0}^{t_K} \left( \left\langle \frac{\partial J}{\partial u}\bigg|_{t}, v \right\rangle + \eta \left(J - \bar{J} \right) \right) dt + \pd{\bar{J}}{s}
\label{e:LSSgradTan}
\end{equation}

This expression can be written as a function of $v'(t)$ from equation \eqref{e:v_proj} by using the following expression

\begin{equation} 
\begin{aligned}
\int_{t_i}^{t_{i+1}} \left\langle \frac{\partial J}{\partial u}\bigg|_{t}, v' \right\rangle \,dt
&= 
\int_{t_i}^{t_{i+1}} \left\langle \frac{\partial J}{\partial u}\bigg|_{t}, v \right\rangle \,dt
- 
\int_{t_i}^{t_{i+1}} \int_{t_i}^{t} \left\langle \frac{\partial J}{\partial u}\bigg|_{t}, f(u(t);s) \right\rangle 
\eta(\tau)\,d\tau\,dt \\
&=
\int_{t_i}^{t_{i+1}} \left\langle \frac{\partial J}{\partial u}\bigg|_{t}, v \right\rangle \,dt
- 
\int_{t_i}^{t_{i+1}} \eta(\tau)
\int_{\tau}^{t_{i+1}} \left\langle \frac{\partial J}{\partial u}\bigg|_{t}, f(u(t);s) \right\rangle \,dt\,d\tau \\
&=
\int_{t_i}^{t_{i+1}} \left\langle \frac{\partial J}{\partial u}\bigg|_{t}, v \right\rangle \,dt
- 
\int_{t_i}^{t_{i+1}} \eta(\tau) (J_{i+1} - J(\tau)) \,d\tau \\
&=
\int_{t_i}^{t_{i+1}} \left(\left\langle \frac{\partial J}{\partial u}\bigg|_{t}, v \right\rangle + \eta\,J\right)dt
- J_{i+1}\,\int_{t_i}^{t_{i+1}} \eta(t) \ dt
\end{aligned} 
\end{equation}

\noindent where $J_{i+1} = J(u(t_{i+1});s)$. Note that 

\begin{align*}
v(t_{i+1}) &= v'(t_{i+1}) + f_{i+1} \int_{t_i}^{t_{i+1}} \eta(t) \ dt, \\
 \langle f_{i+1}, v(t_{i+1}) \rangle &= \langle f_{i+1}, v'(t_{i+1}) \rangle + \| f_{i+1} \|^2_2 \int_{t_i}^{t_{i+1}} \eta(t) \ dt \\
 0 &= \langle f_{i+1}, v'(t_{i+1}) \rangle + \| f_{i+1} \|^2_2 \int_{t_i}^{t_{i+1}} \eta(t) \ dt \\
 \frac{\langle f_{i+1}, v'(t_{i+1}) \rangle}{\| f_{i+1} \|^2_2} &= -\int_{t_i}^{t_{i+1}} \eta(t) \ dt
\end{align*}

\noindent where $f_{i+1} = f(u(t_{i+1});s)$, Therefore

\begin{equation}
\int_{t_i}^{t_{i+1}} \left\langle \frac{\partial J}{\partial u}\bigg|_{t}, v' \right\rangle \,dt = \int_{t_i}^{t_{i+1}} \left(\left\langle \frac{\partial J}{\partial u}\bigg|_{t}, v \right\rangle + \eta\,J\right)dt
+ J_{i+1}\,\frac{\langle f_{i+1}, v'(t_{i+1}) \rangle}{\| f_{i+1} \|^2_2}
\label{e:proj_grad}
\end{equation}

Equation \eqref{e:proj_grad} is substituted into equation  \eqref{e:LSSgradTan} to obtain

\begin{align*}
\frac{d\bar{J}}{ds} &= 
\frac1T \sum_{i=0}^{K-1}
 \int_{t_i}^{t_{i+1}} \left( \left\langle \frac{\partial J}{\partial u}\bigg|_{t}, v' \right\rangle - \eta \bar{J} \right) \ dt - \frac1T \sum_{i=0}^{K-1} \frac{\langle f_{i+1}, v(t_{i+1}) \rangle}{\| f_{i+1} \|^2_2} J_{i+1} + \pd{\bar{J}}{s} \\
\frac{d\bar{J}}{ds} &= 
\frac1T \sum_{i=0}^{K-1}
 \int_{t_i}^{t_{i+1}} \left\langle \frac{\partial J}{\partial u}\bigg|_{t}, v' \right\rangle  \ dt + \frac1T \sum_{i=0}^{K-1} \left( -\bar{J} \int_{t_i}^{t_{i+1}} \eta(t) \ dt - \frac{\langle f_{i+1}, v(t_{i+1}) \rangle}{\| f_{i+1} \|^2_2} J_{i+1} \right)  + \pd{\bar{J}}{s} \\
\end{align*}

\noindent Finally,

\[
\frac{d\bar{J}}{ds} =  
\frac1T \sum_{i=0}^{K-1}
 \int_{t_i}^{t_{i+1}} \left\langle \frac{\partial J}{\partial u}\bigg|_{t}, v' \right\rangle \ dt
+ \frac1T \sum_{i=0}^{K-1} \frac{\langle f_{i+1}, v(t_{i+1}) \rangle}{\| f_{i+1} \|^2_2} \left(\bar{J} - J_{i+1}\right) + \pd{\bar{J}}{s}
\]
where $T \equiv t_K-t_0$.

\section{Adjoint MSS Derivation}
\label{a:AdjMSS}

For a dynamical system with a finite number of states $n$, equation \eqref{e:MSSgrad} can be rewritten using the notation of equation \eqref{e:MSSsys} 

\begin{equation}
\frac{d\bar{J}}{ds} = \ubg^T \ubv + \bh + \pd{\bar{J}}{s}
\label{e:gradVec}
\end{equation} 

\noindent where $\ubg^T = (\bg_1^T, \bg_2^T, ... \bg_K^T, 0)$ is a $1 \times (K+1)n$ vector and 

\begin{equation}
\bg_i^T =  
\frac1T \int_{t_{i-1}}^{t_{i}} \frac{\partial J}{\partial u}\bigg|_{t}^* \phi^{t_{i-1},t}  \ dt
+ \frac1T \left(\bar{J} - J_i\right) \frac{f_{i}|^* \phi^{t_{i-1},t_i}}{\|f_{i}\|^2_2} , 
\label{e:g_def}
\end{equation}

\noindent Where $f_i \equiv f(u(t_i);s)$ and $J_{i} = J(u(t_{i});s)$. The second term in equation \eqref{e:gradVec} is defined as

\begin{align}
\bh &= \sum_{i=1}^K \left[
\frac1T \int_{t_{i-1}}^{t_{i}}  \frac{\partial J}{\partial u}\bigg|_{t}^* \left(\int_{t_i}^{t}  \phi^{\tau,t} \pd{f}{s}\bigg|_{\tau} \ d\tau \right) \ dt
+ \frac1T \frac{(\bar{J} - J_i)}{\|f_i\|_2^2} \int_{t_{i-1}}^{t_{i}} f_i^* \phi^{\tau,t_i} \pd{f}{s}\bigg|_{\tau} \ d\tau \right] \label{e:h_def_1}\\
&= \sum_{i=1}^K \left[
\frac1T \int_{t_{i-1}}^{t_{i}} \pd{f}{s}\bigg|_{\tau}^* \left( \left(\int_{\tau}^{t_{i}}  \phi^{* \ \tau,t} \frac{\partial J}{\partial u}\bigg|_{t} \ dt \right) \ d\tau
+ \frac1T \frac{(\bar{J} - J_i)}{\|f_i\|_2^2} \phi^{* \ \tau,t_i} f_i  \right)\ d\tau \right] \label{e:h_def_2}
\end{align}

\noindent The tangent equation \eqref{e:MSSsys} and the sensitivity equation \eqref{e:gradVec} can be combined as follows

\begin{equation}
\frac{d\bar{J}}{ds}=\left(\begin{array}{c|c}
\ubg^T & 0
\end{array}\right) 
\left(\begin{array}{c}
\ubv \\\hline
\ubw 
\end{array}\right)
+ \bh + \pd{\bar{J}}{s}
+
\left(\begin{array}{c|c}
\uhbv^T & \uhbw^T
\end{array}\right) \left[ \left(\begin{array}{c|c}
-\uuline{I} & \uuline{\bB}^T \\\hline 
\uuline{\bB} & \epsilon \uuline{I} \end{array}\right) \left(\begin{array}{c}
\ubv \\\hline
\ubw 
\end{array}\right) - \left(\begin{array}{c}
0 \\\hline
\ubb
\end{array}\right) \right]
\label{e:MSSadj1}
\end{equation}

\noindent where $\uhbv$ and $\uhbw$ are defined as the adjoint variables. Next, rearrange equation \eqref{e:MSSadj1} as follows 

\begin{equation}
\frac{d\bar{J}}{ds}=-\left(\begin{array}{c|c}
0 & \ubb^T
\end{array}\right) 
\left(\begin{array}{c}
\uhbv \\\hline
\uhbw 
\end{array}\right)
+ \bh + \pd{\bar{J}}{s} 
+
\left(\begin{array}{c|c}
\ubv^T & \ubw^T
\end{array}\right) \left[ \left(\begin{array}{c|c}
-\uuline{I} & \uuline{\bB}^T \\\hline 
\uuline{\bB} & \epsilon \uuline{I} \end{array}\right) \left(\begin{array}{c}
\uhbv \\\hline
\uhbw 
\end{array}\right) + \left(\begin{array}{c}
\ubg \\\hline
0
\end{array}\right) \right] 
\label{e:MSSadj2}
\end{equation}

\noindent One can choose $\uhbv$ and $\uhbw$ to satisfy the following adjoint equation

\begin{equation}
\left(\begin{array}{c|c}
-\uuline{I} & \uuline{\bB}^T \\\hline 
\uuline{\bB} & \epsilon \uuline{I} \end{array}\right) \left(\begin{array}{c}
\uhbv \\\hline
\uhbw 
\end{array}\right) = \left(\begin{array}{c}
-\ubg \\\hline
0
\end{array}\right) 
\label{e:MSSadj3}
\end{equation}

\noindent with Schur complement:

\begin{equation}
(\uuline{\bB}\uuline{\bB}^T + \epsilon \uuline{I}) \ul{\hbw} = - \uuline{\bB} \ubg
\label{e:MSSadjSchur}
\end{equation}

\noindent If $\uhbv$ and $\uhbw$ satisfy equation \eqref{e:MSSadj3}, then the sensitivities can be computed as follows:

\begin{equation}
\frac{d\bar{J}}{ds}= -\ubb^T \uhbw + \bh + \pd{\bar{J}}{s} 
\label{e:MSSadjGrad1}
\end{equation}

\noindent Using equations \eqref{e:b_def} and \eqref{e:h_def_2}, the definitions of $\ubb$ and $\bh$, respectively, equation \eqref{e:MSSadjGrad1} can be rewritten as:

\begin{equation}
\frac{d\bar{J}}{ds} = \sum_{i=1}^K \int_{t_{i-1}}^{t_i} \pd{f}{s}\bigg|_{\tau}^T \left[   \phi^{* \ \tau,t_i } P_{t_i} \hat{\bf w}_i  + \frac1T \left(\int_{\tau}^{t_{i}} \phi^{* \ \tau,t} \frac{\partial J}{\partial u}\bigg|_{t} \ dt \right) 
+ \frac1T \frac{\bar{J} - J_i}{f_i^T f_i} \phi^{* \ \tau,t_i} f_i  \right] \ d\tau
\label{e:MSSadjGrad2}
\end{equation}

\noindent From \eqref{e:MSSadj3}: 

\begin{gather}
 -\hbv_{i-1} -\hbw_{i-1} + \Phi_i^T \hbw_i = -\bg_i \quad i = 0,1,...,K \label{e:mss_adj}\\
 \hbw_0 = 0, \hbw_K = -\hbv_K \nonumber\\
 \Phi_i \hbv_{i-1} - \hbv_i + \epsilon \hbw_i = 0
\end{gather}

\noindent For equations \eqref{e:mss_adj} and \eqref{e:g_def} to be true, the adjoint solution $\hat{w}(t)$ in time segment $i-1$ should be

\[
 \hat{w}(t) = \phi^{* \ \tau,t_i } P_{t_i} \hat{\bf w}_i  + \frac1T \left(\int_{t}^{t_{i}} \phi^{* \ t,\tau} \frac{\partial J}{\partial u}\bigg|_{\tau} \ d\tau \right) 
+ \frac1T \frac{\bar{J} - J_i}{f_i^T f_i} \phi^{* \ \tau,t_i} f_i
\]

\noindent With this choice of $\hw(t)$, equation \eqref{e:MSSadjGrad2} simplifies to:

\begin{equation}
\frac{d\bar{J}}{ds} = \sum_{i=1}^K \int_{t_{i-1}}^{t_i} \pd{f}{s}\bigg|_{t}^T \hat{w}(t) \ d\tau
\label{e:MSSadjGrad4}
\end{equation}

\noindent Sensitivities of $\bar{J}$ with respect to many parameters $s$ can be computed with a single solution of $\hat{w}(t)$. 

\section{Tangent MSS Algorithm}
\label{a:tanMSSalg}

Since the adjoint MSS Schur complement \eqref{e:MSSadjSchur1} is identical to the tangent MSS Schur complement \eqref{e:MSSschur1} the adjoint and tangent algorithms are very similar:

\begin{framed}

\noindent\textbf{Tangent MSS Solver} \\
{\it Inputs:} Initial condition for the governing equations $u_0$, Spin-up time $t_0$, Specified time horizon and checkpoints $t_0,t_1,...,t_K$, Initial guess for $Kn \times 1$ vector $\ubw$, which contains the Lagrange multipliers at checkpoints $1$ to $K$ (default value 0); \\
{\it Ouputs:} Sensitivity $d\bar{J}/ds$ \\
{\it Calls:} MATVEC algorithm that computes $\ubR = (\uuline{\bB}\uuline{\bB}^T + \epsilon \uuline{I}) \ubw - \beta\ubb$ where $\ubR$ is a $Kn \times 1$ residual vector. \\ 
\begin{enumerate}
\item Time integrate the governing equations \eqref{e:dyn_sys} to compute $u(t)$ for the specified time horizon.  
\item To form the right hand side of the linear system, $\ubb$, use the MATVEC algorithm with $\beta = -1$ and $\ubw=0$. 
\item Use some iterative algorithm to solve equation \eqref{e:MSSschur1}. To compute the left hand side $(\uuline{\bB}\uuline{\bB}^T + \epsilon \uuline{I})\ubw$, use MATVEC with $\beta = 0$.  
\item Compute the sensitivity $d\bar{J}/ds$ using equation \eqref{e:MSSgrad}.  
\end{enumerate}

\end{framed}

Next, a serial MATVEC algorithm is presented. Note that $t_i^-$ and $t_i^+$ refer to the time at checkpoint $i$ in time segments $i-1$ and $i$, respectively. For example, $v(t_i^-)$ is the tangent solution at checkpoint $i$ computed in time segment $i-1$. \\

\begin{framed}
\noindent\textbf{Serial Tangent MATVEC Algorithm} \\
{\it Inputs:} $\ubw$, a $Kn \times 1$ vector of the Lagrange multipliers at checkpoints 1 to $K$; $\beta$, a scalar; \\
{\it Ouputs:} $\ubR$, a $Kn \times 1$ residual vector \\
MATVEC computes $\ubR = (\uuline{\bB}\uuline{\bB}^T + \epsilon \uuline{I}) \ubw - \beta\ubb$ \\
\begin{enumerate}
\item For all time segments, compute $w(t_{i-1}^+)$ by integrating
$\frac{dw}{dt} = -\left(\pd{f}{u}\right)^* w,\;
t\in(t_{i-1},t_i)$ backwards in time with the terminal condition $w(t_i) = P_{t_i}\bw_i$. 
\item Save $\bv_{i-1} \equiv w(t_{i-1}^+) - \bw_{i-1}$. If $i=1$, save $\bv_0 \equiv \hw(t_0^+)$. 
\item For all time segments, compute $v'(t_i^-)$ by integrating
$\frac{dv'}{dt} = \pd{f}{u} v' + \beta\pd{f}{s},\; t\in(t_{i-1},t_{i}]$
with the initial condition $v'(t_{i-1}) = \bv_{i-1}$. 
\item For all time segments, compute $\bR_i = P_{t_i}\,v'(t_i^-) - \bv_i + \epsilon \bw_i$.  If $i=K$, $\bv_K = - \bw_K$. 
\end{enumerate}
\end{framed}

\section{Estimating the largest eigenvalue of the MSS KKT Schur complement}
\label{a:MSS_max_ev}



The largest eigenvalue of the MSS KKT Schur complement, $\mu_{max}$, is related to the positive Lyapunov exponent discussed in section \ref{ss:ivp}. This can be shown with the singular value decomposition (SVD) of the tangent transition matrix $\Phi_i$ defined in equation \eqref{e:MSS0fwd},

\begin{gather}
\Phi_i = \bU_i \Sigma_i \bV_i^T=\left(\begin{array}{cccc}
\vertbar & \vertbar & & \vertbar \\
\bU_i^1 & \bU_i^2 & \hdots & \bU_i^n \\
\vertbar & \vertbar & & \vertbar 
\end{array}\right) \left(\begin{array}{cccc}
\sigma_i^1 & & & \\
& \sigma_i^2 & & \\
& & \ddots & \\
& & & \sigma_i^n
\end{array}\right) \left(\begin{array}{ccc}
\horzbar & [\bV_i^1]^T & \horzbar  \\
\horzbar & [\bV_i^2]^T & \horzbar  \\
 & \vdots & \\
\horzbar & [\bV_i^n]^T & \horzbar  \\
\end{array}\right) \label{e:transitionSVD}
\end{gather}

\noindent where $\bU_i$ and $\bV_i$ are orthonormal matrices and $\Sigma_i$ is a diagonal matrix with singular values $\sigma_i^j$ along the main diagonal. The singular values are indexed in descending order of magnitude, that is $\sigma_i^1 > \sigma_i^2 > \hdots > \sigma_i^n$. 


 Interestingly the singular value $\sigma_i^1$ can be interpreted as a finite time approximation of the largest Lyapunov exponent. The maximum Lyapunov exponent can be defined as \cite{Temam:1997:dynamics}
 
 \begin{equation}
 \Lambda^{max} = \max_{j} \lim_{t \to \infty} \sup \frac1t \text{ln}\left(\sigma^j(\phi^{0,t}) \right)
 \label{e:Lya_def}
 \end{equation}
 
 \noindent where $\sigma^j(\phi^{0,t})$ is the $j$th singular value of a the tangent propagator $\phi^{0,t}$, which is a $n \times n$ matrix for a system with $n$ states. Recall that $\Phi_i = P_{t_i}\phi^{t_{i-1},t_i}$. The projection operator $P_{t_i}$ only removes any component parallel to $f(u)$, which happens to be the Lyapunov covariant vector for the zero Lyapunov exponent. Therefore, $P_{t_i}$ has no impact on the covariant vector corresponding to $\Lambda_{max}$ and 
 

 \[
 \Lambda^{max} = \max_{j} \lim_{\Delta T_i \to \infty} \sup \frac{1}{\Delta T_i} \text{ln}\left(\sigma^j(\Phi_i) \right)
 \]
 
 \noindent where $\Delta T_i=t_i-t_{i-1}$ is the time segment length for segment $i$. Since $\sigma_i^1$ is defined as the maximum singular value of $\Phi_i$ in equation \eqref{e:transitionSVD},
 
 \begin{equation}
 \Lambda^{max} = \lim_{\Delta T_i \to \infty} \tilde{\Lambda}^{max}_i, \qquad 
 \tilde{\Lambda}_i = \frac{1}{\Delta T_i} \text{ln}\left(\sigma^1_i \right)
 \label{e:FTLya_def}
 \end{equation}
 
\noindent The quantity $\tilde{\Lambda}^{max}_i$ can be interpreted as a finite time approximation of $\Lambda^{max}$ in time segment $i$. From equation \eqref{e:FTLya_def}, the singular value $\sigma^1_i$ can be written in terms of the finite time Lyapunov exponent 

\begin{equation}
\sigma^1_i = e^{\tilde{\Lambda}^{max}_i \Delta t_i }
\label{e:lyapunov_svd_link}
\end{equation}

Recall from section \ref{ss:mss_mu_max} that $\tilde{\Lambda}^{max}_i \Delta t_i$ can be much larger than 1 due to variations in $\tilde{\Lambda}^{max}_i$ over each time segment. Because of this, it can be assumed that 

 \begin{equation}
 \max_i[\sigma_i^1] \gg 1
 \label{e:assum0}
 \end{equation}
 
\noindent if $K$ or $\delta t_i$ is sufficiently large. 

Next, the connection between $\mu_{max}$ and the $\sigma^1_i$ is considered. Recall from equations \eqref{e:MSSschur1} and \eqref{e:MSS0schur2} that the MSS Schur complement matrix $\uuline{\bA}$ is a block-tridiagonal matrix with $\Phi_i\Phi_i^T$ the main block diagonal and $\Phi_i$ and $\Phi_i^T$. Therefore, the eigenvalues of $\uuline{\bA}$ must be related to the singular values $\sigma_i^j$ of $\Phi_i$. It can be shown that the largest eigenvalue $\mu_{max}$ of the $\uuline{\bA}$ is 

\begin{equation}
\mu_{max} \approx 1 + \epsilon + \max_i [\sigma_i^1]^2
\label{e:max_eig_sigma}
\end{equation}

To show this, consider the case where the maximum singular value is $\sigma_l$, corresponding the tangent propagator $\Phi_l$. Then, for $\ubx = (0,\vdots,0,[\bU_l^1]^T,0,\vdots,0)^T$, 

\begin{align}
\uuline{\bA} \ubx &= \left(\begin{array}{cccc}
\Phi_1 \Phi_1^T + (1+\epsilon) I & -\Phi_2^T & & \\
-\Phi_2 & \Phi_2 \Phi_2^T + (1+\epsilon) I & -\Phi_3^T & \\
 & \ddots & \ddots & \ddots  \\
 & & -\Phi_K & \Phi_K\Phi_K^T + (1+\epsilon)I 
\end{array}\right) \left(\begin{array}{c}
0 \\
\vdots \\
0 \\
\bU_l^1 \\
0 \\
\vdots \\
0
\end{array}\right) \\
&= \left(\begin{array}{c}
0 \\
\vdots \\
-\Phi_l^T \bU_l^1 \\
(\Phi_l\Phi_l^T + (1+\epsilon)I) \bU_l^1 \\
-\Phi_{l+1} \bU_l^1 \\
\vdots \\
0
\end{array}\right) = \left(\begin{array}{c}
0 \\
\vdots \\
-\bV_l \Sigma_l \bU_l^T \bU_l^1 \\
(\bU_l \Sigma_l^2 \bU_l + (1+\epsilon)I) \bU_l^1 \\
-\bU_{l+1} \Sigma_{l+1} \bV_{l+1}^T \bU_l^1 \\
\vdots \\
0
\end{array}\right) \\
&= \left(\begin{array}{c}
0 \\
\vdots \\
-\sigma_l^1 \bV_l^1 \\
([\sigma_l^1]^2 + 1 + \epsilon) \bU_l^1 \\
-\bU_{l+1} \Sigma_{l+1} \bV_{l+1}^T \bU_l^1 \\
\vdots \\
0
\end{array}\right)
\label{e:MSS_max_evec}
\end{align}

\noindent

Rows $l-1$ and $l+1$ of $\uuline{A}\ubx$ are negligible relative to row $i$ if $\sigma_l^1 \gg 1$ as in equation \eqref{e:assum0} and $\epsilon \ge 0$.

In row $l-1$, since by definition $\|\bU_l^1\|_2 = \| \bV_l^1 \|_2 = 1$,

 \begin{equation}
 |([\sigma_l^1]^2 + 1 + \epsilon) \bU_l^1| \gg |\sigma_l^1 \bV_l^1|
 \label{e:assum1}
 \end{equation}
 
 \noindent The magnitude of row $l+1$ of $\uuline{A}\ubx$ is a maximum when $\bU_1^l = \bV_{l+1}^1$, as $\bV_{l+1}^1$ will be stretched by the largest singular value for time segment $l+1$, $\sigma_{l+1}^1$, so
 
 \[
 \bU_{l+1} \Sigma_{l+1} \bV_{l+1}^T \bV_{l+1}^1 = \sigma_{l+1}^1 \bU_{l+1} \ge \bU_{l+1} \Sigma_{l+1} \bV_{l+1}^T \bU_l^1
 \]
 
 \noindent since $\sigma_l^1$ is the largest singular value for all time segments, $\sigma_l^1 > \sigma_{l+1}^1$ and since by definition $\|\bU_l^1\|_2 = \| \bU_{l+1}^1 \|_2 = 1$
 
 \begin{equation}
 |([\sigma_l^1]^2 + 1 + \epsilon) \bU_l^1| \gg |\sigma_{l+1}^1 \bU_{l+1}| \ge |\bU_{l+1} \Sigma_{l+1} \bV_{l+1}^T \bU_l^1|
 \label{e:assum2}
 \end{equation}
 
\noindent The inequalities in equations \eqref{e:assum1} and \eqref{e:assum2} show that
 
 \begin{equation}
 \uuline{\bA} \ubx \approx \left(\begin{array}{c}
0 \\
\vdots \\
0 \\
([\sigma_l^1]^2 + 1 + \epsilon) \bU_l^1 \\
0 \\
\vdots \\
0
\end{array}\right) = ([\sigma_l^1]^2 + 1 + \epsilon) \ubx
 \end{equation}
 
\noindent Therefore, the largest eigenvalue of $\uuline{A}$ is 
 
 \begin{equation}
 \mu_{max} \approx [\sigma_l^1]^2 + 1 + \epsilon
 \label{e:max_eig_sigma2}
 \end{equation}
 
 \noindent with the corresponding eigenvector $\ubx$. Since $\sigma_l^1$ is by definition the largest singular value for all time segments, equations \eqref{e:max_eig_sigma2} and \eqref{e:max_eig_sigma} are equivalent. 
 
 Note that $\mu_{max} \gg 1$ in the typical spectrum shown in figure \ref{f:MSSspectrum}. Since this spectrum was computed for $\epsilon = 0$, the assumption $\sigma_l^1 \gg 1$ is consistent with a typical spectrum of $\uuline{\bA}$.

\noindent Therefore, 

\begin{equation}
\mu_{max} \approx 1 + \epsilon + \max_i e^{2\tilde{\Lambda}^{max}_i \Delta T_i}
\label{e:max_eig_lya1}
\end{equation}

\section*{Acknowledgments}
Research for this paper was conducted with Government support under FA9550-11-C-0028 and awarded by the Department of Defense, Air Force Office of Scientific Research, National Defense Science and Engineering Graduate (NDSEG) Fellowship, 32 CFR 168a. 

\bibliographystyle{elsarticle-num}
\section*{Bibliography}
\bibliography{main}

\newcommand{\noopsort}[1]{} \newcommand{\printfirst}[2]{#1}
  \newcommand{\singleletter}[1]{#1} \newcommand{\switchargs}[2]{#2#1}
\begin{thebibliography}{10}
\expandafter\ifx\csname url\endcsname\relax
  \def\url#1{\texttt{#1}}\fi
\expandafter\ifx\csname urlprefix\endcsname\relax\def\urlprefix{URL }\fi
\expandafter\ifx\csname href\endcsname\relax
  \def\href#1#2{#2} \def\path#1{#1}\fi

\bibitem{Giles:2000:adj}
M.~Giles, N.~Pierce, An introduction to the adjoint approach to design, Flow,
  Turbulence and Combustion 65 (2000) 393--415.

\bibitem{Jameson:1988:adj}
A.~Jameson, Aerodynamic design via control theory, Journal of Scientific
  Computing 3~(3) (1988) 233--260.

\bibitem{Reuther:2001:adjAC}
J.~Reuther, A.~Jameson, J.~J. Alonso, M.~J. Rimlinger, D.~Sanders, Constrained
  multipoint aerodynamic shape optimization using an adjoint formulation and
  parallel computers, Journal of Aircraft 36~(1) (1999) 51--74.

\bibitem{Martins:2005:adjAS}
J.~R. R.~A. Martins, J.~J. Alonso, J.~J. Reuther, A coupled-adjoint sensitivity
  analysis method for high-fidelity aero-structural design, Optimization and
  Engineering 6~(1) (2005) 33--62.
\newblock {http://dx.doi.org/DOI:10.1023/B:OPTE.0000048536.47956.62}

\bibitem{Darmofal:2002:adapt}
D.~Venditti, D.~Darmofal, Grid adaptation for functional outputs: Application
  to two-dimensional inviscid flow, Journal of Computational Physics 176 (2002)
  40--69.

\bibitem{Giles:2002:adjEE}
M.~Giles, E.~S{\"u}li, Adjoint methods for pdes: a posteriori error analysis
  and postprocessing by duality, Acta Numerica 11 (2002) 145--236.

\bibitem{Gunzburger:2002:PFC:640624}
M.~Gunzburger, Perspectives in Flow Control and Optimization, Society for
  Industrial and Applied Mathematics, Philadelphia, PA, USA, 2002.

\bibitem{Wang:2009:Thesis}
Q.~Wang, Uncertainty quantification for unsteady fluid flow using adjoint-based
  approaches, {PhD} dissertation, Stanford University (2009).

\bibitem{Wang:2013:hyper}
Q.~Wang, K.~Duraisamy, J.~Alonso, G.~Iaccarino, Risk assessment of scramjet
  unstart using adjoint-based sampling methods, AIAA Journal 50~(3) (2012)
  581--592.

\bibitem{Wang:2014:LSS2}
Q.~Wang, R.~Hui, P.~Blonigan, Least squares shadowing sensitivity analysis of
  chaotic limit cycle oscillations, Journal of Computational Physics 267 (2014)
  210--224.

\bibitem{Wang:2013:LSS_PF}
Q.~Wang, S.~Gomez, P.~Blonigan, A.~Gregory, E.~Qian, Towards scalable
  parallel-in-time turbulent flow simulations, Physics of Fluids 25.
\newblock {http://dx.doi.org/10.1063/1.4819390}.

\bibitem{Blonigan:2014:KS}
P.~Blonigan, Q.~Wang, Least squares shadowing sensitivity analysis of a
  modified {K}uramoto--{S}ivashinsky equation, Chaos, Solitons, and Fractals 64
  (2014) 16--25.

\bibitem{Blonigan:2014:MG}
P.~Blonigan, Q.~Wang, Multigrid-in-time for sensitivity analysis of chaotic
  dynamical systems, Numerical Linear Algebra with Applications 21~(2).

\bibitem{Blonigan:2016:AIAA}
P.~Blonigan, Q.~Wang, E.~Nielsen, B.~Diskin, Least squares shadowing
  sensitivity analysis of chaotic flow around a two-dimensional airfoil, AIAA
  2016-0296, 2016.

\bibitem{gomez:2013:masters}
S.~Gomez, Parallel multigrid for large-scale least squares sensitivity,
  Master's thesis, Massachusetts Institute of Technology, Cambridge, MA (2013).

\bibitem{Hasselblatt:2002:hyperbolic}
B.~Hasselblatt, Hyperbolic dynamical systems, in: Handbook of Dynamical Systems
  1A, Elsevier, North Holland, 2002, pp. 239--319.

\bibitem{Ruelle:1997:SRB}
D.~Ruelle, Differentiation of {SRB} states, Communications in Mathematical
  Physics 187 (1997) 227--241.

\bibitem{Kuznetsov:2009:plykin}
S.~Kuznetsov, A non-autonomous flow system with {P}lykin type attractor,
  Communications in Nonlinear Science and Numerical Simulation 11~(9) (2009)
  3487--3491.

\bibitem{Eyink:2004:ensmbl}
G.~Eyink, T.~Haine, D.~Lea, Ruelle's linear response formula, ensemble adjoint
  schemes and {L}\'evy flights, Nonlinearity 17~(5) (2004) 1867--1889.

\bibitem{Bonatti:2005:Hyper}
C.~Bonatti, L.~Diaz, M.~Viana, Uniform Hyperbolicity: A Global Geometric and
  Probabilistic Perspective, Springer, 2005.

\bibitem{Lorenz:1963:det}
E.~Lorenz, Deterministic nonperiodic flow, Journal of the Atmospheric Sciences
  20 (1963) 130--141.

\bibitem{Lea:2000:climate_sens}
D.~Lea, M.~Allen, T.~Haine, Sensitivity analysis of the climate of a chaotic
  system, Tellus 52A (2000) 523--532.

\bibitem{Gallavotti:1995:chaosHyp1}
G.~Gallavotti, E.~Cohen, Dynamical ensembles in stationary states, Journal of
  Statistical Physics 80 (1995) 931--970.

\bibitem{Gallavotti:1995:chaosHyp2}
G.~Gallavotti, E.~Cohen, Dynamical ensembles in nonequilibrium statistical
  mechanics, Physical Review Letters 74 (1995) 2694--2697.

\bibitem{Albers:2006:StrucStab}
D.~Albers, J.~Sprott, Structural stability and hyperbolicity violation in
  high-dimensional dynamical systems, Nonlinearity 19 (2006) 1801--1847.

\bibitem{Kim:1986:dns_channel}
J.~Kim, P.~Moin, R.~Moser, Turbulence statistics in fully developed channel
  flow at low {R}eynolds number, Journal of Fluid Mechanics 177 (1986)
  133--166.

\bibitem{Medic:2012:CTR}
G.~Medic, J.~Joo, S.~Lele, O.~Sharma, Prediction of heat transfer in a turbine
  cascade with high levels of free-stream turbulence, in: Proceedings of the
  2012 Summer Program, Center for Turbulence Research, Stanford, 2012, pp.
  147--155.

\bibitem{Bose:2010:expLES}
S.~Bose, P.~Moin, D.~You, Grid-independent large-eddy simulation using explicit
  filtering, Physics of Fluids 22 (2010) 105103.

\bibitem{Strogatz:1994:chaos}
S.~Strogatz, Nonlinear Dynamics and Chaos: With Applications to Physics,
  Biology, Chemistry, and Engineering, Westview Press, Philadelphia, PA, 1994.

\bibitem{Lea:2002:ocean}
D.~Lea, T.~Haine, M.~Allen, J.~Hansen, Sensitivity analysis of the climate of a
  chaotic ocean circulation model, Journal of the Royal Meteorological Society
  128 (2002) 2587--2605.

\bibitem{Hicken:2014:chaosOpt}
A.~Ashley, J.~Hicken, Low {R}eynolds number numerical solutions of chaotic
  flow, in: AIAA Aviation 2014 Symposium on the Theory of Computing, Atlanta,
  Georgia, United States, 2014, aIAA-2014-2434.

\bibitem{Thuburn:2005:FP}
J.~Thuburn, Climate sensitivities via a {F}okker-{P}lanck adjoint approach,
  Quarterly Journal of the Royal Meteorological Society 131~(605) (2005)
  73--93.

\bibitem{Blonigan:2014:density}
P.~Blonigan, Q.~Wang, Probability density adjoint for sensitivity analysis of
  the mean of chaos, Journal of Computational Physics 270 (2014) 660--686.

\bibitem{Nyquist:1928:therEq}
H.~Nyquist,Thermal agitation of electric charge in conductors. Phys. Rev. 32 (1928) 110--113.
\newblock {http://dx.doi.org/10.1103/PhysRev.32.110}.

\bibitem{Kubo:1966:FDT}
R.~Kubo, The
  fluctuation-dissipation theorem, Reports on Progress in Physics 29~(1)
  (1966) 255.

\bibitem{leith:1975:climate}
C.~Leith, Climate response and fluctuation dissipation, Journal of the
  Atmospheric Sciences 32~(10) (1975) 2022--2026.

\bibitem{majda:2005:information}
A.~Majda, R.~Abramov, M.~Grote,
  Information Theory and Stochastics for Multiscale Nonlinear Systems, CRM Monograph Series, American Mathematical Society, 2005.

\bibitem{Abramov:2008:FDT}
R.~Abramov, A.~Majda, New approximations and tests of linear
  fluctuation-response for chaotic nonlinear forced-dissipative dynamical
  systems, Journal of Nonlinear Science 18 (2008) 303--341.

\bibitem{Abramov:2007:FDTblend}
R.~Abramov, A.~Majda,
  Blended response
  algorithms for linear fluctuation-dissipation for complex nonlinear dynamical
  systems, Nonlinearity 20~(12) (2007) 2793.

\bibitem{youngSRB}
L.-S. Young,What are {SRB}
  measures, and which dynamical systems have them?, Journal of Statistical
  Physics 108~(5-6) (2002) 733--754.
\newblock {http://dx.doi.org/10.1023/A:1019762724717}.

\bibitem{Benettin:1980:Lyapunov}
G.Benettin, L.~Galgani, A.~Giorgilli, J.~Strelcyn, Lyapunov characteristic
  exponents for smooth dynamical systems and for {H}amiltonian systems; a
  method for computing all of them. part 2: Numerical application, Meccanica
  15~(1) (1980) 21--30.

\bibitem{Doedel:1989:orbits}
E.~J. Doedel, M.~J. Friedman, Numerical computation of heteroclinic orbits,
  Journal of Computational and Applied Mathematics 26~(1-2) (1989) 155--170.

\bibitem{Pilyugin:1999:shadow}
S.~Y. Pilyugin, Shadowing in dynamical systems, 1st Edition, Vol. 1706 of
  Lecture Notes in Mathematics, Springer-Verlag, New York, 1999.

\bibitem{Wang:2014:LSSthm}
Q.~Wang, Convergence of the least squares shadowing method for computing
  derivative of ergodic averages, SIAM Journal of Numerical Analysis 52~(1)
  (2014) 156--170.

\bibitem{Paige:1975:minres}
C.~C. Paige, M.~A. Saunders, Solution of sparse indefinite systems of linear
  equations, SIAM Journal of Numerical Analysis 12 (1975) 617--629.

\bibitem{Golub1996}
G.~H. Golub, C.~F.~V. Loan, Matrix Computations, The Johns Hopkins Univ. Press,
  Baltimore, 1996.

\bibitem{Sanchez:2010:MS}
J.~Sanchez, M.~Net, On the multiple shooting continuation of periodic orbits by
  {N}ewton-{K}rylov methods, International Journal of Bifurcation and Chaos
  20~(1) (2010) 43--61.

\bibitem{Friedhoff:2013:MG}
S.~Friedhoff, R.~Falgout, T.~Kolev, S.~MacLachlan, J.~Schroder, A
  multigrid-in-time algorithm for solving evolution equations in parallel, in:
  Sixteenth Copper Mountain Conference on Multigrid Methods, Copper Mountain,
  Colorado, 2013.

\bibitem{Falgout:2015:mgrit}
R.~D. Falgout, S.~Friedhoff, T.~V. Kolev, S.~P. MacLachlan, J.~B. Schroder,
  Parallel time integration with multigrid, SIAM Journal on Scientific
  Computing 36~(6) (2015) 635--661.

\bibitem{Falgout:2015:mgritCFD}
R.~D. Falgout, A.~Katz, T.~V. Kolev, J.~B. Schroder, A.~Wissink, U.~M. Yang,
  Parallel time integration with multigrid reduction for a compressible fluid
  dynamics application, j. Comp. Phys., (submitted). LLNL-JRNL-663416. (2015).

\bibitem{Saad:1986:GMRES}
Y.~Saad, M.~H. Schultz, Gmres: A generalized minimum residual method for
  solving nonsymmetric linear systems, SIAM Journal on Scientific and
  Statistical Computing 7~(3) (1986) 856--869.

\bibitem{Dowell:1982:chaos_plate}
E.~Dowell, Flutter of a buckled plate as an example of chaotic motion of a
  deterministic autonomous system, Journal of Sound and Vibration 85~(3) (1982)
  333--344.

\bibitem{Dowell1975}
E.~H. Dowell, Aeroelasticity of Plates and Shells, Noordhoff, Leyden, The
  Netherlands, 1975.

\bibitem{Dowell:1966:AIAA_plate}
E.~Dowell, Nonlinear oscillations of a fluttering plate, part i, AIAA Journal 4
  (1966) 1267--1275.

\bibitem{Dowell:1967:AIAA_plate}
E.~Dowell, Nonlinear oscillations of a fluttering plate, part ii, AIAA Journal
  5 (1967) 1856--1862.

\bibitem{Yang:2011:Thesis}
S.~Yang, A shape hessian-based analysis of roughness effects on fluid flows,
  {PhD} dissertation, University of Texas at Austin (2011).

\bibitem{Fritsch:1980:pchip}
F.~N. Fritsch, R.~E. Carlson, Monotone piecewise cubic interpolation, SIAM
  Journal of Numerical Analysis 17~(2) (1980) 238--246.

\bibitem{Ruelle:1985:erg}
J.-P. Eckmann, D.~Ruelle, Ergodic theory of chaos and strange attractors,
  Reviews of Modern Physics 57~(3) (1985) 617--656.

\bibitem{Hyman:1986:KS}
J.~M. Hyman, B.~Nicolaenko, The {K}uramoto-{S}ivashinsky equation: A bridge
  between {PDE}'s and dynamical systems, Physica D: Nonlinear Phenomena 18:1-3
  (1986) 113--126.

\bibitem{Kuramoto:1976:reaction}
Y.~Kuramoto, T.~Tsuzuki, Persistent propagation of concentration waves in
  dissipative media far from thermal equilibrium, Prog. Theor. Phys. 55 (1976)
  356--369.

\bibitem{Kuramoto:1978:reaction}
Y.~Kuramoto, Diffusion-induced chaos in reaction systems, Suppl. Prog. Theor.
  Phys. 64 (1978) 364--367.

\bibitem{Sivashinsky:1977:flames1}
G.~Sivashinsky, Nonlinear analysis of hydrodynamic instability in laminar
  flames, part i. derivation of basic equations, Acta Astronautica 4 (1977)
  1177--1206.

\bibitem{Sivashinsky:1977:flames2}
G.~Sivashinsky, Nonlinear analysis of hydrodynamic instability in laminar
  flames, part ii. numerical experiments, Acta Astronautica 4 (1977)
  1207--1221.

\bibitem{Sivashinsky:1980:film}
G.~Sivashinsky, D.~Michelson, On irregular wavy flow of a liquid film down a
  vertical plane, Progr. Theoret. Phys. 63 (1980) 2112--2114.

\bibitem{Sapsis:2013:locDim}
T.~Sapsis, Attractor local dimensionality, nonlinear energy transfers and
  finite-time instabilities in unstable dynamical systems with applications to
  two-dimensional fluid flows, Proceedings of the Royal Society A 469 (2013)
  20120550.

\bibitem{Farmer:1991:shadow}
J.~D. Farmer, J.~J. Sidorowich, Optimal shadowing and noise reduction, Physica
  D 47 (1991) 373--392.

\bibitem{Temam:1997:dynamics}
R.~Temam, Infinite-Dimensional Systems in Mechanics and Physics, 2nd Edition,
  Vol.~68 of Applied Mathematical Sciences, Springer-Verlag, New York, 1997.

\end{thebibliography}

\end{document}